\documentclass[sn-mathphys-num,iicol]{sn-jnl}

\usepackage{graphicx}%
\usepackage{caption}
\usepackage{subcaption}
\usepackage{lipsum}%
\usepackage{multirow}%
\usepackage{amsmath,amssymb,amsfonts}%
\usepackage{amsthm}%
\usepackage{mathrsfs}%
\usepackage[title]{appendix}%
\usepackage{xcolor}%
\usepackage{cuted}
\usepackage{textcomp}%
\usepackage{manyfoot}%
\usepackage{booktabs}%
\usepackage{algorithm}%
\usepackage{algorithmicx}%
\usepackage{algpseudocode}%
\usepackage{listings}%
\usepackage{makecell}
\usepackage{mathtools}
\usepackage{bbm}
\usepackage{cleveref}
\usepackage{rotating}
\usepackage{mhchem}
\usepackage{xurl}
\usepackage[normalem]{ulem}
\theoremstyle{thmstyleone}%
\theoremstyle{thmstyletwo}%

\theoremstyle{thmstylethree}%

\begin{document}
\title{Machine Learning-Informed 3+1 Sterile Neutrino Global Fits using Posterior Density Estimation of Electron Disappearance Data}


\author*[1]{\fnm{Joshua} \sur{Villarreal} \orcid{0000-0001-9690-1310}}\email{villaj@mit.edu}

\author[1]{\fnm{Julia} \sur{Woodward} \orcid{0009-0006-1636-7562}}\email{julia785@mit.edu}

\author[1]{\fnm{John} \sur{Hardin} \orcid{0000-0001-8871-8065}}\email{jmhardin@mit.edu}

\author[1]{\fnm{Janet} \sur{Conrad} \orcid{0000-0002-6393-0438}}\email{conrad@mit.edu}

\affil[1]{\orgdiv{Department of Physics}, \orgname{Massachusetts Institute of Technology}, \orgaddress{\street{77 Massachusetts Avenue}, \city{Cambridge}, \postcode{02138}, \state{MA}, \country{USA}}}

\abstract{Global analyses of particle physics data are integral for validating and scrutinizing published results of experiments.  Global fits of anomalous oscillation data which search for one or more eV-scale sterile neutrinos are particularly challenging both to evaluate and to reconcile in the global picture. Fits (especially joint ones) to oscillation data suffer from significant computational burdens, such as likelihood intractability, making traditional Markov Chain-Monte Carlo all but impossible. Given evidence both supporting and challenging beyond Standard Model physics across neutrino experiments of various baselines, energies, and detection techniques, the global search for sterile neutrinos requires additional tools in order to determine whether sterile neutrinos remain a viable solution to unexplained anomalies. Furthermore, both a Bayesian and frequentist interpretation of sterile neutrino data is needed for a complete assessment of longstanding tensions in the field. Techniques from the machine learning subfield of simulation-based inference have a natural application to such a problem. In this contribution, we illustrate some of the outstanding questions of the global picture of light sterile neutrinos by focusing on experiments searching with the disappearance of electron (anti)neutrinos, and look to posterior density estimation strategies to craft answers, including comparisons to a machine-learning-based frequentist approach.}


\maketitle

\section{Introduction}\label{sec:introduction}

Within the field of particle physics, progress often comes from analyzing disparate data sets to extract subtle signals. Searches for light, eV-scale sterile neutrinos are a prime example of this ``global fit’’ approach \cite{Diaz:2019fwt, Hardin:2022muu, Dasgupta:2021ies, Acero:2022wqg, Giunti:2022btk}. These analyses combine short-baseline data from neutrino experiments with different baselines, energies, and detection technologies, each with unique systematic uncertainties and sensitivities, to identify the most interesting parameter space for future searches and to assess the probability that anomalous signals are due to specific new physics models. Most often, the models add a single non-interacting, hence ``sterile’’ neutrino to the three neutrinos that interact via the Standard Model’s weak force; hence this class is called 3+1 models.

Within 3+1 studies, and in the case of many global fits to particle physics data, fitters most often rely on frequentist statistical tests of compatibility of the data. In the frequentist paradigm, a global fit typically proceeds by evaluating a test statistic over a physical model’s parameter space and interpreting its profile by quoting a \textit{confidence level} (CL). The standard recipe profiles the likelihood ratio between a given hypothesis and the best-fit case, assuming that the resulting test statistic follows a $\chi^2$ distribution as described by Wilks’ theorem \cite{Wilks1938-yc}. While this approximation enables rapid computation of CLs, it relies on a set of assumptions that are not met by searches for small oscillatory signals (such as 3+1) because effective degrees of freedom vary over the parameter space and systematic uncertainties can deviate from Gaussianity \cite{Algeri2020-cw}. In these situations, the nominal coverage of the inferred CLs can fail consistency checks under repeated pseudoexperiments \cite{Hardin:2022qdh}, motivating alternative approaches to inference for complex or non-Gaussian global analyses. A very common example of this is the setup of trials-based tests first described by Feldman and Cousins \cite{feldman-cousins}. However, while full Feldman–Cousins constructions of CLs are exact, they are computationally infeasible to compute at the scale of modern sterile-neutrino data. In previous work \cite{10.1088/2632-2153/ae040c}, we introduced a trials-based framework that uses machine learning (in particular, simulation-based inference [SBI]) to construct well-calibrated CLs without relying on Wilks’ theorem. That work demonstrates that frequentist global fits for sterile-neutrino scenarios can be made both principled and tractable.

This paper provides the Bayesian counterpart to our frequentist method~\cite{10.1088/2632-2153/ae040c}. As our example, we focus on electron-flavor disappearance data in a 3+1 framework; formulating the global analysis by once again leveraging recent developments in SBI to use the same simulator to efficiently generate a large number of pseudoexperiments over a variety of physics hypothesis, and perform density estimation to directly and efficiently estimate the posterior distribution over the parameters. In this work, we compare several state-of-the-art posterior density estimation strategies, based on discrete and continuous normalizing flows and diffusion models, and the extracted posterior from the neural likelihood ratio estimation strategy utilized in Ref.~\cite{10.1088/2632-2153/ae040c}.  We then use these learned posteriors to draw parameter constraints and credibility regions (the Bayesian counterpart to frequentist CLs) from the published reactor-, source-, and accelerator-based $\nu_e/\overline{\nu}_e$ disappearance data. The result is a Bayesian analysis said to be ``likelihood free", and does not rely on Markov Chain Monte Carlo sampling methods to gather posterior estimates leading to significant improvements in run time at inference.

The Bayesian treatment of sterile neutrino data is an important companion to the frequentist version of the analysis. The statistical questions asked by each analysis are not redundant, but complementary methods of model fitting and uncertainty quantification. Answers to both analysis philosophies are informative in a field where results from different experiments can be in tension; in recent times, both fits are commonly reported (e.g., Refs.~\cite{icecube-sterile-prl, icecube-sterile-prd}). 

Also, the simulator-level approach lets us propagate non-Gaussian and experiment-specific systematics by construction. In the Bayesian setting, these effects can be marginalized hierarchically and reported transparently via full posterior structure (including multi-modality and parameter degeneracies). Credible regions (the highest posterior density [HPD]) and posterior predictive checks provide diagnostics that complement frequentist coverage studies.

The two approaches do not represent double the effort. The paradigms can share the same pseudoexperiment surrogates and common algorithmic pipelines. In particular, amortized posteriors enable fast maximum-\emph{a posteriori} (MAP) estimates, which can satisfy the ranking criterion of the best-fit point estimates required by Feldman–Cousins-like test statistics \cite{feldman-cousins, 10.1088/2632-2153/ae040c}. As we show, some posterior estimators (notably \texttt{SNPE-C} in our benchmarks) deliver MAPs with low error at very low latency, making them attractive plug-and-play upgrades for trials-based approaches like Ref.~\cite{10.1088/2632-2153/ae040c}.

The methods we recommend in this work can be applicable across many types of global (or even single-experiment) fits within particle physics. In this work, we use the specific example of electron-flavor disappearance in a 3+1 model: a relatively simple case that, nonetheless, tests a physical model dependent on more than one parameter and relies on three qualitatively different data sets taken at reactors (which infer oscillations of $\overline{\nu}_e$), or radioactive sources and accelerators (oscillations of $\nu_e$). These experiments combine data from particles ($\nu_e$) and antiparticles ($\overline{\nu}_e$) in experiments with complementary baselines, energies, and detector designs, leading to very different systematic uncertainties, emblematic of the issues that global fitters of any particle physics data sets will face.

The goal of the analysis is to determine the best pair of parameters of the 3+1 model for electron neutrino disappearance found by estimating the posterior distribution and constructing credible regions over a grid. Our method trains density estimation strategies on data generated by the simulation code described in Ref.~\cite{10.1088/2632-2153/ae040c}, which produces pseudoexperiments capturing experiments' reported systematic and statistical uncertainties. We then evaluate calibration and accuracy using withheld test-set samples, and use the trained model to fit the published data to obtain HPD regions of credibility that can be  compared (qualitatively) with the confidence regions computed from a frequentist framework.

Taken together, the frequentist and Bayesian analyses present a coherent, simulation-driven view of the electron-disappearance sector in 3+1 models. Agreement between HPD regions and trials-based confidence regions corroborates that the conclusions of the global fit are not artifacts. Differences between results can help localize sensitivity to priors, non-Gaussian systematics, ordering rules, or even inconsistencies in the simulator itself. The result is a more robust global picture.

The remainder of this paper is structured as follows. Section~\ref{sec:aboutdata} briefly reviews the data sets and the 3+1 model used in the example global fit.   Section~\ref{sec:sbi-methods} details the simulator construction, training data preparation, and the various posterior estimation strategies evaluated. Section~\ref{sec:results} presents the benchmark results for each estimator and applies the chosen method to the $\nu_e$ and $\overline{\nu}_e$ disappearance data. Section ~\ref{sec:upgrading-freq} details the integration of the posterior estimators with the frequentist framework of Ref. ~\cite{10.1088/2632-2153/ae040c}. Section ~\ref{sec:discussion} discusses the implications of the Bayesian fits and their limitations.   Finally, Section~\ref{sec:conclusion} summarizes our conclusions.

\section{About Electron Neutrino Disappearance\label{sec:aboutdata}}
\begin{sidewaystable}[tp!]
\caption{Relevant parameters of experiments used in this study. Characteristic $L$ and $E$ values are approximate, see reference (column 2) for more details. Column 5 provides the reported anomaly significance of a 3+1 signal in numbers of standard deviations (or NR if the best fit point is not reported), though each provides a best fit point regardless of 3+1 preference; for these, see Appendix \ref{app:datadists}.}\label{tab:experiments}
\begin{tabular*}{\textwidth}{@{\extracolsep{\fill}}l c c c c l}
\toprule
Name & Ref. & $L$ [m] & $E$ [MeV] & Significance & \multicolumn{1}{c}{Description} \\
\midrule
\multicolumn{6}{@{}p{\dimexpr\textwidth-2\tabcolsep\relax}@{}}{\centering\textbf{Reactor Experiments ($\bar{\nu}_e$)}}\\
STEREO      & \cite{stereocollaboration2020antineutrino} & 9--11    & 1.625--7.125 & NR        & ILL research reactor, extended detector \\
PROSPECT    & \cite{prospectResults} & 7--10    & 0.8--7.2     & NR        & ORNL research reactor, extended detector \\
DANSS       & \cite{Skrobova2023} & 10--12   & 1--7         & NR & Kalinin power reactor, moving baseline system \\
NEOS/RENO   & \cite{PhysRevD.105.L111101} & 224/294  & 1--10        & NR        & Hanbit power reactor, two detectors with RENO setting SM flux \\
\midrule
\multicolumn{6}{@{}p{\dimexpr\textwidth-2\tabcolsep\relax}@{}}{\centering\textbf{Source Experiments ($\nu_e$)}}\\
SAGE \& GALLEX & \cite{Gavrin2013,KOSTENSALO2019542,Giunti:2010zu, SAGE:2009eeu, PhysRevC.73.045805} & 0--0.7 \& 0--1.9 & 0.426, 0.747, 0.752, 0.811, 0.813 & 2.3$\sigma$ & $^{51}$Cr and $^{37}$Ar sources; single segment $^{71}$Ga target \\
BEST          &  \cite{PhysRevLett.128.232501, PhysRevD.105.L051703} & 0--1.7   & 0.426, 0.747, 0.752 & $>$5$\sigma$ & $^{51}$Cr source; two-segment $^{71}$Ga target \\
\midrule
\multicolumn{6}{@{}p{\dimexpr\textwidth-2\tabcolsep\relax}@{}}{\centering\textbf{Accelerator Experiments ($\nu_e$)}}\\
KARMEN/LSND & \cite{Conrad:2011ce} & $17.7/29.8$ & 17.3--52.8 & NR & $\nu_e$-carbon cross section measurements from pion decay-at-rest $\nu_e$ \\
\botrule
\end{tabular*}
\end{sidewaystable}

Throughout this paper we will use the 3+1 model for electron flavor neutrino disappearance as our example, although our methods have application to most global fits performed in particle physics.  This section provides a very brief introduction to the model we will use and the data sets that will be fit.    

The 3+1 model describes potential neutrino oscillations between the three standard light flavors and a sterile neutrino. Our example focuses on whether a neutrino produced as  electron-flavored ($\nu_e$) will interact downstream from the production point as a $\nu_e$ or whether it will become non-interacting, hence sterile ($\nu_s$). The probability that a produced $\nu_e$ continues to interact as a $\nu_e$ downstream has a oscillatory dependence, given by:
\begin{equation}
P_{\nu_e \rightarrow \nu_e} = 1-4|U_{e4}|^2 (1-|U_{e4}|^2) \sin^2(\Delta_{41}L/E).~ \label{PUe4}
\end{equation} 
where $L$ [m], the distance traveled (often called the ``baseline'') and $E$ [MeV] the energy of the neutrino, are experimentally chosen parameters.    From this formula, one can extract two underlying parameters of the 3+1 model: $|U_{e4}|$, associated with the amplitude of the oscillation; and $\Delta_{41} \equiv 1.27 \Delta m_{41}^2$,  the frequency of the oscillations in 
$\text{eV}^2$.  
In principle, to keep $P_{\nu_e \to \nu_e} \in [0, 1]$, we require $|U_{e4}| \in [0, 1]$; however, in our studies we will use $|U_{e4}| \leq 1 / \sqrt{2}$ which turns out to fully explores the electron-flavor disappearance parameter space.  In a 3+1 model, the formula for the antineutrino disappearance probability is identical to that of neutrinos (Eq.~\ref{PUe4}).

Table~\ref{tab:experiments} summarizes the experiments used in our electron-flavor global fit. Published data for each are reproduced in the appendix (see Fig.~\ref{fig:published}).

\subsection{Reactor Experiments \label{reactorfluxdiscussion}}
The reactor experiments with the most extensive reach in 3+1 parameter space are STEREO \cite{stereocollaboration2020antineutrino}, PROSPECT \cite{prospectResults}, DANSS \cite{Skrobova2023}, and NEOS \cite{PhysRevD.105.L111101}. Reactors emit anti-electron neutrinos ($\bar{\nu}_e$) detected via inverse beta decay interactions $\overline{\nu}_e (p, n) e^+$, allowing precise reconstruction of baseline $L$ and antineutrino energy $E$. To detect sterile neutrino oscillations, the measured event rate as a function of $L/E$ is compared to Standard Model (SM) predictions for the event rate given the calculated reactor antineutrino flux.  Evidence for a sterile neutrino is a deficit of measured $\overline{\nu}_e$ compared to flux expectations, hence a reduction in rate.  The SM-based reactor flux prediction used in this paper is the Huber-Mueller (HM) flux \cite{Huber:2011wv}, and is a common choice.   However, this flux prediction is known to have two unresolved issues when compared with data. 

The first issue is that the HM flux over-predicts the total rate of neutrino interactions, an effect called the Reactor Antineutrino Anomaly (RAA) that was initially reported in 2011~\cite{Mention:2011rk}. Since that time, multiple new flux models addressing this issue, summarized in Ref.~\cite{Giunti:2021kab}, have been put forward. These models range in data-to-prediction ratios from $0.925_{-0.023}^{+0.025}$ for the HKSS model to $0.975_{-0.021}^{+0.022}$ for the KI model, but there is no consensus ``best choice'' for among these for a new flux prediction~\cite{Giunti:2021kab}. In principle, we can assign a theoretical uncertainty to the HM flux and proceed with our fits. In practice, however, SM explanations for this overall deficit are degenerate with a signature for sterile neutrino oscillations in certain parameter space. Examining Eq.~\ref{PUe4}, one sees that as $\Delta m^2$ becomes large relative to the $L/E$ of a given experiment, oscillations become rapid. Due to experimental resolution, the oscillatory behavior becomes unobservable and the term depending on $\Delta m^2$ averages to $1/2$, consistent with an overall deficit. To remove the confusion between an SM or sterile neutrino source of a deficit and to reduce the theoretical uncertainty from the flux models, we remove the overall normalization mismatch to perform ``shape-only'' fits that demand oscillatory behavior.  This reflects the view that neutrino oscillations should explicitly be seen to oscillate, but this conservative choice does diminish sensitivity to high values of $\Delta m_{41}^2$ ($\gtrapprox \mathcal{O}(10\,\text{eV}^2)$).

The second data-to-simulation comparison issue for the HM reactor flux is a deviation in the predicted reactor flux at energies $\approx 5$~MeV.  Known as the ``$5$ MeV bump", the $\sim 10\%$ excess is seen in all modern reactor data sets across baselines from $\sim 1$ m to 1000 m.  The source of the effect is not understood, but it is not due to 3+1 oscillations, since that would produce a deficit rather than an excess and also would show $L$ dependence (see Eq.~\ref{PUe4}). First-principles models based on reactor fuel $^{235}$U and $^{239}$Pu consumption aimed at explaining the excess have not successfully fit all of the reactor data, even when introducing an empirical parameter ~\cite{DayaBay:2025ngb}.   In order to address this issue, short baseline (SBL) reactor experiments employ two types of mitigation.   The first, used by DANSS and NEOS/RENO, fits ratios of data sets across the same $E$ range but at different $L$-values, which will remove the ``bump effect''.   DANSS took ratios of data from the same detector at different locations.  NEOS and RENO are experiments located at a short (NEOS) and long (RENO) baseline at the same reactor complex, and in the discussion below are always treated as an ``experimental pair" \cite{PhysRevD.105.L111101}.    The second, used by STEREO and PROSPECT, fits for the $L$ dependence in each energy bin separately.    Due to these mitigations, the reactor results that are used in the fits in this paper have negligible bias from the $5$~MeV bump.

\subsection{Source Experiments}
The SAGE \cite{Gavrin2013, SAGE:2009eeu}, GALLEX \cite{Giunti:2010zu}  and BEST \cite{PhysRevLett.128.232501, PhysRevD.105.L051703} experiments use fluxes produced by radioactive sources with $\text{MCi}$-level decay rates. These sources emit monoenergetic electron neutrinos ($\nu_e$) produced through electron capture in $^{51}$Cr and $^{37}$Ar sources, and are paired with gallium-filled detectors. The $\nu_e$ interactions convert gallium nuclei to germanium ($\nu_e (\ce{^{71}Ga}, \ce{^{71}Ge}) e^-$), which is periodically removed and counted to compare against well-predicted rates \cite{Elliott:2023xkb}. While low normalization uncertainty allows high sensitivity to non-zero $|U_{e4}|$, poor segmentation limits $\Delta m_{41}^2$ sensitivity, restricting measurements to $\Delta m_{41}^2 \gtrapprox \mathcal{O} (10\,\text{eV}^2)$. Thus, the reactor and source data sets are complementary in multiple ways.

\subsection{Accelerator-based Decay-at-rest Experiments}

The third class of experiments for our example fits uses fluxes produced at two accelerator-based decay-at-rest sources. These sources impinge $800\,\text{MeV}$ protons on a target, producing pions that stop and produce a decay-at-rest chain resulting in 4$\pi$ emission of $\nu_\mu$, $\overline{\nu}_\mu$, and $\nu_e$ in equal fractions with a well-understood energy spectrum and a normalization known to the $\sim 10\%$ level. The highly reproducible nature of the decay-at-rest sources allows cross comparison of widely-separated experiments, even at  facilities on separate continents.    
This study compares data from the KARMEN detector \cite{Zeitnitz1993-ll} at Rutherford Lab in the UK to the LSND detector \cite{ATHANASSOPOULOS1997149} at LAMPF in the US.  The experiments were located at different $L$ from their respective decay-at-rest sources:  KARMEN is at baseline of $17.7\,\text{m}$ and LSND is at baseline of $29.8\,\text{m}$. Thus, comparison of the relative rate of $\nu_e$ interactions in the two experiments, published in the form of energy-dependent cross-sections, allows a search for the $L$ dependent effects of $\nu_e$ 3+1 oscillations.    In the global fit presented here, we use the comparison study presented in  Ref.~\cite{Conrad:2011ce}. In the fits and discussion below, like the NEOS/RENO ratio, KARMEN and LSND will be treated as an inseparable experimental pair.

\section{Training, Evaluation, and Selection of Posterior Density Estimator}\label{sec:sbi-methods}

\subsection{Training data preparation\label{subsec:data-prep}}

The Monte Carlo simulation used in this work, \texttt{sblmc}, is described in detail in Refs.~\cite{Diaz:2019fwt, Hardin:2022muu}, and is the same as used in the frequentist SBI global fit approach paper in Ref.~\cite{10.1088/2632-2153/ae040c}. In short, the codebase is designed for calculations of model log-likelihoods based on experimental data, requiring either a covariance matrix or a set of pull parameters provided by experiments during result releases. In addition to a set of model expectations, the code can be used to generate a set of realizations for any sterile neutrino model for any combination of experiments.

To generate the training- and test-set data used in this study, we define an evenly spaced $50 \times 50$ grid in $\log_{10} U_{e4}$, $\log_{10} \Delta m_{41}^2$ for $U_{e4} \in [0.001, 1/\sqrt{2}]$ and $\Delta m_{41}^2 \in [0.01, 100]\,\text{eV}^2$. For each of the experiments listed in Tab.~\ref{tab:experiments}, we generate 1000 mock realizations per parameter point resulting in 2,500,000 total pseudoexperiments.

\subsection{Posterior Estimation Strategies Considered Here\label{subsec:posterior-estimators}}

To estimate the posterior density, we must begin with the following:
\begin{itemize}
\item Model parameters $\theta$, which, in this contribution, are shorthand for the sterile oscillation parameters $\theta \equiv (U_{e4}, \Delta m_{41}^2)$ in Eq.~\ref{PUe4},
\item A pre-defined prior distribution $\mathbb{P} (\theta)$, and 
\item A simulator capable of sampling experimental data $\mathbf{x} \sim \mathbb{P} (\mathbf{x} | \theta)$.
\end{itemize}
Given these inputs, the strategy is to infer, for observed experimental data $\mathbf{x}_\text{obs}$, the posterior probability density $\mathbb{P} (\theta | \mathbf{x}_\text{obs})$. 

Conditional density estimators approximate the posterior distribution with neural networks. That is, the true posterior $\mathbb{P} (\theta | \mathbf{x}) \approx q_{\psi} (\theta)$, with parameters $\psi \approx F(\mathbf{x} | \phi)$ for $F$ the neural network having parameters $\phi$ trained with loss criterion $\mathcal{L} (\phi) = - \sum_i \log q_{F (\mathbf{x}_i | \phi) } (\theta_i)$. In the case where the neural network $F$ is sufficiently expressive and the size of the training data is large enough, as $\mathcal{L}$ is minimized, $q_{F (\mathbf{x}_i | \phi)} \to \mathbb{P} (\theta | \mathbf{x}_i)$. Such a conditional density estimator can then be used to approximate the posterior probability density as $\mathbb{P} (\theta | \mathbf{x}_\text{obs} ) \approx q_{F (\mathbf{x}_\text{obs} | \phi)} (\theta)$. Presently, neural networks called \textit{normalizing flows} which learn a sequence of invertible transformations between two probability distributions are used for such tasks and have shown promise in a variety of applications \cite{Cranmer2020}. 

All of the strategies except for the direct amortized neural likelihood ratio estimation were investigated using implementations provided by the \texttt{sbi} Python package \cite{BoeltsDeistler_sbi_2025}, itself a heavy user of \texttt{PyTorch} \cite{Ansel_PyTorch_2_Faster_2024}.  The method described in Sec.~\ref{subsubsec:dnre} was implemented using \texttt{tensorflow} \cite{tensorflow2015-whitepaper}, 

\subsubsection{Sequential Neural Posterior Estimation\label{subsubsec:snpe_c}}

For a complete overview, refer to Ref.~\cite{snpe-c}. In \textit{sequential neural posterior estimation}, a conditional density estimator is refined across multiple rounds of training to approximate the true posterior distribution $\mathbb{P} (\theta | \mathbf{x}_\text{obs})$. In the first round, parameters $\theta \sim \mathbb{P} (\theta)$ are drawn from the prior and used to generate simulated data $\mathbf{x}$. The neural network $F (\mathbf{x} | \phi)$ is then trained such that the induced density $q_F (\theta)$ approximates the posterior distribution $\mathbb{P} (\theta | \mathbf{x})$. In subsequent rounds of training, parameters are sampled instead from a proposal distribution $\tilde{\mathbb{P}} (\theta)$ which is meant to concentrate probability mass in regions more informative about observed data $\mathbf{x}_\text{obs}$. 

However, training on these samples can bias the estimator. In particular, when $\theta \sim \tilde{\mathbb{P}(\theta)}$ in the second step of the training sequence, minimizing the loss function $\mathcal{L} (\phi) = - \sum_i \log q_{F(\mathbf{x}_i | \phi)} (\theta_i)$ over these samples forces $q_F (\theta) \to \mathbb{P} (\theta|x) \frac{\tilde{\mathbb{P}}(\theta) \mathbb{P}(\mathbf{x})}{\mathbb{P}(\theta) \tilde{\mathbb{P}}(\mathbf{x})}$ (for $\tilde{\mathbb{P}} (\mathbf{x}) = \int_\theta \tilde{\mathbb{P}}(\theta) \mathbb{P} (\mathbf{x}|\theta)$). \textit{Automatic posterior transformation} (which, based on precedent, is known as \texttt{SNPE-C}, indicating the third generation of previously-developed \textit{sequential neural posterior estimation} strategies) corrects for this bias by modifying the loss function, training the network to minimize $\tilde{\mathcal{L}}(\phi) = - \sum_i \log [q_{F(\mathbf{x} | \phi)} \tilde{\mathbb{P}}(\theta) / \mathbb{P}(\theta)]$. Over multiple rounds of training, the network weights $\phi$ are modified by selecting simulations from sequentially updated proposals $\tilde{\mathbb{P}}(\theta)$. As the network sends $q_{F(\mathbf{x} | \phi)} (\theta) \to \mathbb{P} (\theta | \mathbf{x})$, $\mathbb{E} [\tilde{\mathcal{L}}]$ is minimized, agnostic to the proposal $\tilde{\mathbb{P}} (\theta)$. Therefore, an advantage of \texttt{SNPE-C} is that the network can leverage data from multiple training rounds simultaneously by simply summing their respective losses.  With some additional technical details (to which we refer the reader to Ref.~\cite{snpe-c}), one can use normalizing flows (parameterized by $\psi$) for the prior, proposal, and posterior estimates, taking advantage of advances in machine learning.

\subsubsection{Flow Matching\label{subsubsec:fm}}

\textit{Flow matching posterior estimation} (\texttt{FMPE}) replaces the discrete normalizing flow in the preceding section with a continuous normalizing flow, where the transformation from a simple base distribution to the posterior distribution is defined by the solution of an ordinary differential equation. First introduced in Refs.~\cite{lipman2023flowmatchinggenerativemodeling, NEURIPS2023_3663ae53}, we describe the general idea below for completeness.

A \textit{continuous normalizing flow}, parameterized by a continuous ``time" variable $t \in [0, 1]$, much like a discrete normalizing flow, also maps a base to a target distribution. For base distribution $q_0$ and target distribution $q$, and for invertible mapping $\psi_x : \mathbb{R}^n \to \mathbb{R}^n$, the a velocity field $v_{t, x}$ governs this transformation with an ordinary differential equation:
\begin{equation*}
    \label{eq:fmpe-ode}
    \frac{d}{dt} \psi_{t, x} (\theta) = v_{t, x} (\psi_{t, x} (\theta)); \, \psi_{0, x} (\theta) = \theta.
\end{equation*} The final density can by found by integration:
\begin{equation*}
    q (\theta | x) = q_0 (\theta) \exp \bigg{(} - \int_{0}^1 \textrm{div}\,v_{t, x} (\theta_t) dt \bigg{)}.
\end{equation*}

The goal of flow matching is to use a neural network to approximate the velocity field $v_{t, x}$. To ensure the target probability path is found, a well-constructed loss function does the trick; we refer the reader to Refs.~\cite{lipman2023flowmatchinggenerativemodeling, NEURIPS2023_3663ae53} for the details. The key insight is that discrete normalizing flows depend on a prespecified density estimator architecture, whereas continuous normalizing flows can be more expressive by learning arbitrary probability paths.

\subsubsection{Neural Posterior Score Estimation\label{subsubsec:npse}}

\textit{Neural Posterior Score Estimation} (\texttt{NPSE}) formulates simulation-based inference as a conditional diffusion process that learns to approximate the score of the posterior distribution \cite{pmlr-v202-geffner23a, sharrock2024sequentialneuralscoreestimation}. Details are listed here. \texttt{NPSE} introduces a forward diffusion process that gradually perturbs $\theta$ toward a tractable reference distribution (e.g., a standard Gaussian) according to the stochastic differential equation (SDE)
\begin{equation}
    d\theta_t = f(\theta_t, t)\,dt + g(t)\,dw_t,
    \label{eq:npse-sde}
\end{equation}
where $f$ and $g$ are drift and diffusion coefficients, and $(w_t)_{t \ge 0}$ is a standard Brownian motion. The reverse-time process defining the generative model evolves as
{\small
\begin{align}
d\overline{\theta}_t
&= [-f(\overline{\theta}_t, T-t)
+ g^2(T-t)\nabla_\theta \log p_{T-t}(\overline{\theta}_t | x)] dt \nonumber \\
&\quad + g(T-t)\,dw_t.
\label{eq:npse-reverse-sde}
\end{align}
}

Under this formulation, generating posterior samples corresponds to simulating the reverse diffusion process conditioned on the observation $x_\mathrm{obs}$. Since the true score $\nabla_\theta \log p_t(\theta | x)$ is intractable, \texttt{NPSE} trains a neural network $s_\psi(\theta, x, t)$ to approximate it by minimizing a weighted Fisher divergence, with the internal expectation approximated via Monte Carlo sampling from the prior and the forward diffusion process.

For deterministic inference or density evaluation, \texttt{NPSE} can alternatively employ the probability flow ODE associated with Eq.~\eqref{eq:npse-sde},
\begin{equation}
    \frac{d\theta_t}{dt} = f(\theta_t, t) - \tfrac{1}{2} g^2(t)\nabla_\theta \log p_t(\theta | x),
    \label{eq:npse-ode}
\end{equation}
which shares the same marginal distributions as the stochastic process. Once the score network is trained, posterior samples are drawn by simulating the reverse-time SDE or its deterministic counterpart, initialized from the reference distribution. This provides a flexible and likelihood-free approach to posterior inference with strong generative capabilities.

\subsubsection{Direct Amortized Neural Likelihood Ratio Estimation\label{subsubsec:dnre}}
\texttt{SNPE-C}, \texttt{NPSE}, and FM are learning tasks in which the goal is to find a neural network estimate for the posterior distribution $\mathbb{P}(\theta | \mathbf{x})$. For most practical applications, the true posterior distribution is unknown, so optimizing a machine learning model to estimate the posterior is known as an unsupervised learning task. On the other hand, supervised learning tasks require knowledge of the true labels or target values at training time (not necessarily at the inference step), and tend to be more straightforward.

\begin{figure*}[t!]
\centering

% Left column: tall subfigure
\begin{subfigure}[t!]{0.48\textwidth}
    \centering
    \includegraphics[scale=0.75]{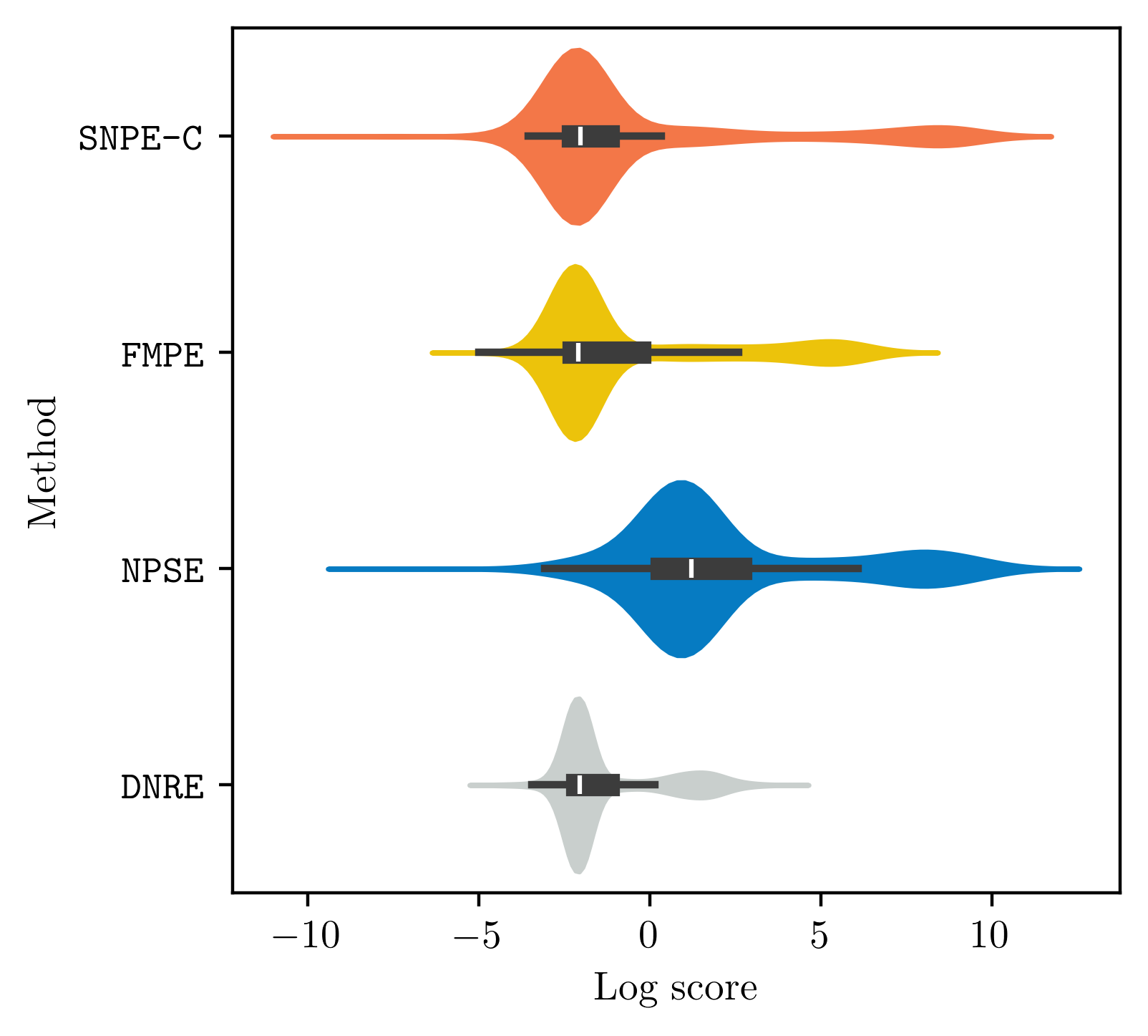}
    \caption{Distribution of log-scores for each flow-based posterior estimator. 
    While the mean log-score is presented as a summary statistic in Tab.~\ref{tab:posterior-estimation-results}, 
    it does not effectively capture the substructure in the metric distributions; each log-score distribution is bimodal, 
    representing the dichotomy of pseudoexperiments to be generated for signal- and null-like physics parameters. 
    The bump at lower values of log-score corresponds to the flow assigning an approximately uniform probability 
    over parameter space when the pseudoexperiment is thrown from a set of physics parameters to which the global fit 
    is not sensitive (indistinguishable from null). 
    The bump at higher values of log-score correspond to the networks assigning a high posterior probability at the true 
    injected parameters when the global fit can distinguish those parameters meaningfully (a signal-like sample). 
    The high log score for null-like samples in the case of the \texttt{NPSE} posterior estimator is physically unreasonable.}
    \label{fig:log-scores}
\end{subfigure}
\hfill
% Right column: two stacked subfigures
\begin{subfigure}[t!]{0.48\textwidth}
    \centering
    % Wrap the stacked plots in a minipage
    \begin{minipage}[t]{\linewidth}
        \centering
        \includegraphics[scale=0.75]{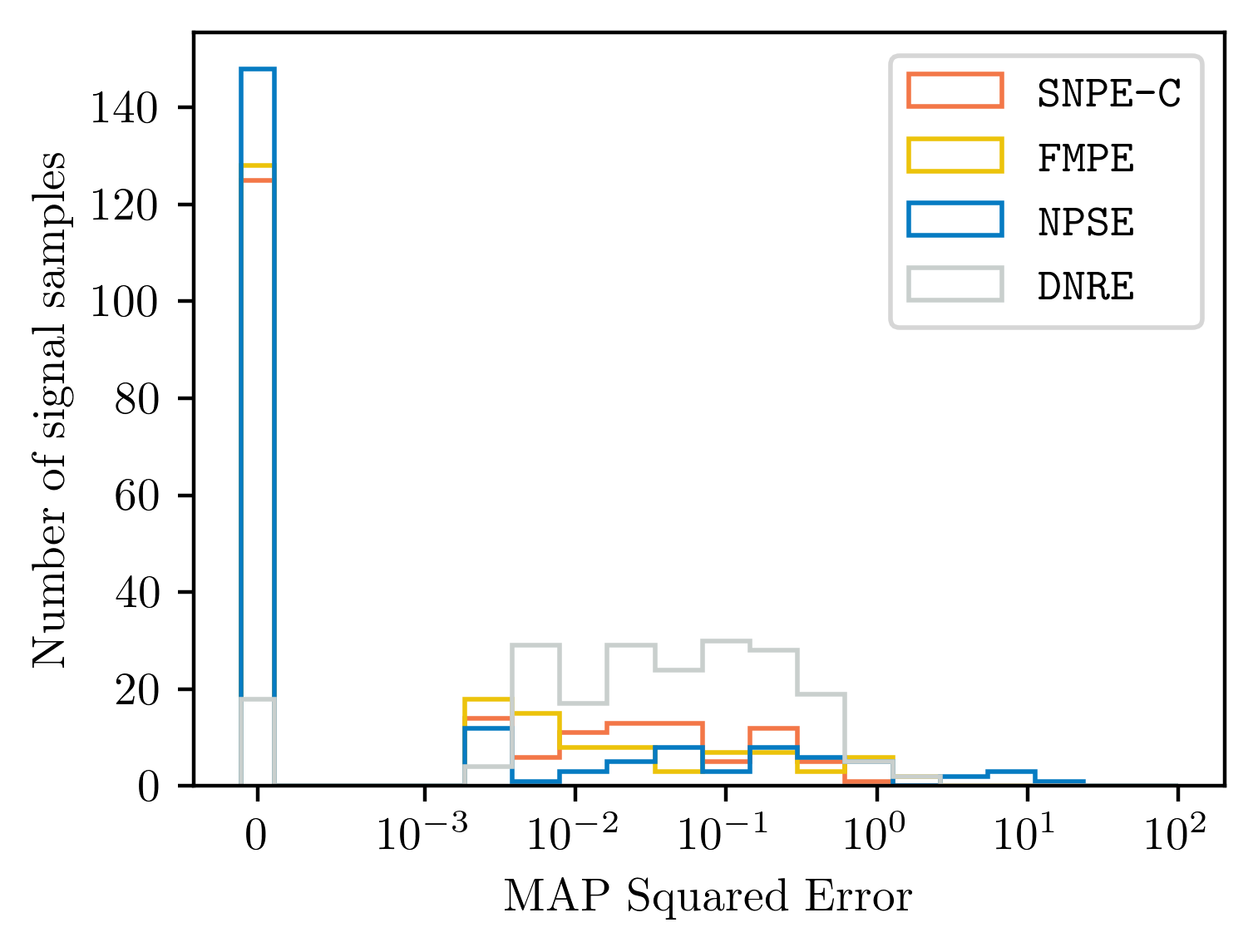}
        \caption{Maximum \textit{a posteriori} (MAP) squared errors on out-of-sample test entries with $\log_{10} U_{e4} > -0.7$, where we expect the global fit to be able to identify a signal.}
        \label{fig:map-errors}
    \end{minipage}

    \vspace{1em}

    \begin{minipage}[t]{\linewidth}
        \centering
        \includegraphics[scale=0.75]{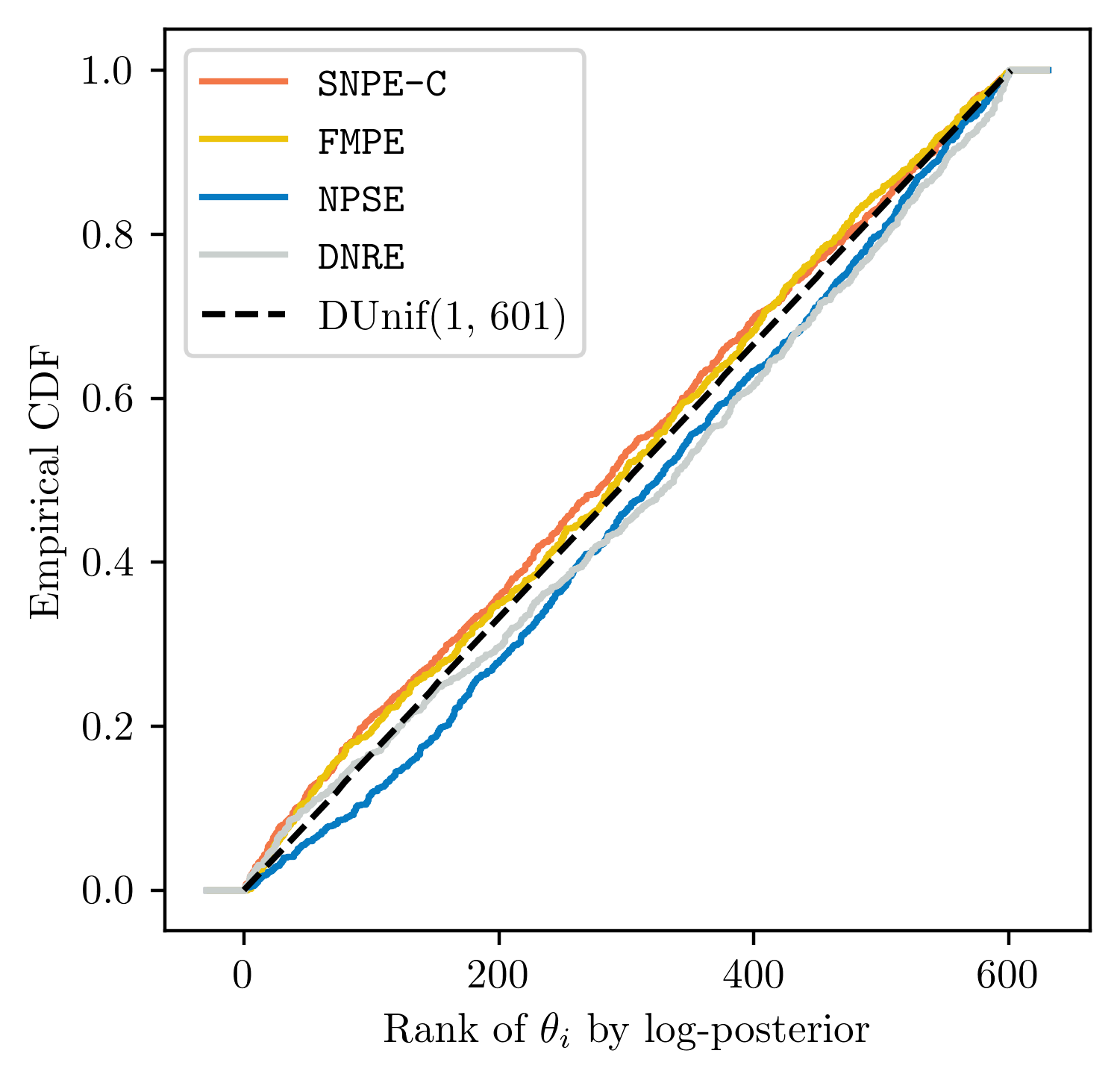}
        \caption{Empirical cumulative distribution of rank scores for each posterior estimator and the \texttt{DNRE} baseline.}
        \label{fig:rank-scores}
    \end{minipage}
\end{subfigure}

\caption{Comparison of the posterior density estimation methods based on flows and the neural likelihood ratio estimator from Ref.~\cite{10.1088/2632-2153/ae040c} using (a) log-score distributions, (b) MAP error statistics, and (c) rank score distributions.}
\label{fig:evaluation-summary}
\end{figure*}

One such strategy, Direct Amortized Neural Likelihood Ratio Estimation (\texttt{DNRE}) \cite{10.1609/aaai.v38i18.30018}, has been used to conduct global fits of anomalous muon neutrino disappearance in a frequentist setting \cite{10.1088/2632-2153/ae040c}. The details of implementation are described therein, but are summarized briefly for completeness here. Samples $\{ (\mathbf{x}_i, \theta_i) \}_{i=1}^N$ are collected via the method of Sec.~\ref{subsec:data-prep}, and in addition, $M$ additional prior-sampled model parameters are collected: $\{ \theta'_j \}_{j=1}^N$. The training set is constructed by two steps of matrix augmentation: samples $\{ (\mathbf{x}_i, \theta_i, \theta'_i) \}_{i=1}^N$ are assigned the label $1$, while samples $\{ (\mathbf{x}_i, \theta'_i, \theta_i) \}_{i=1}^N$ are assigned the label $0$. A classifier $\mathbf{d}_\phi (\mathbf{x}, \theta, \theta')$ is trained using binary cross-entropy loss to predict the correct label given $(\mathbf{x}, \theta, \theta')$; its goal is to determine whether the experimental (Monte Carlo) data was generated from $\theta$ or $\theta'$. The network's output can be transformed into an estimate of the likelihood ratio,

\begin{align*}
    \frac{\mathbb{P} (\mathbf{x} | \theta)}{\mathbb{P} (\mathbf{x}|\theta')} \approx \hat{r} (\mathbf{x} | \theta, \theta') =  
    \frac{\mathbf{d}_\phi (\mathbf{x}, \theta, \theta')}
    {1 - \mathbf{d}_\phi (\mathbf{x}, \theta, \theta')},
\end{align*}

\noindent its intended use case. However, Ref.~\cite{10.1609/aaai.v38i18.30018} points out that the classifier's output can also be used as an estimate of the posterior probability density:
\begin{align}
\label{eq:dnre-logposterior}
\log \mathbb{P} (\theta | \mathbf{x})
&\approx - \mathrm{LogSumExp} \!\left[ - \log \hat{r} (\mathbf{x} | \theta, \theta_i') \right] \notag \\
&\quad + \log M + \log \mathbb{P} (\theta).
\end{align}

\begin{table*}[t]
\caption{Performance of posterior estimation strategies presented in Sec.~\ref{sec:sbi-methods}, including scores for network calibration and timing considerations. Five-fold cross validation was performed to ensure result consistency among disparate validation sets (reported as mean $\pm$ standard deviation). }\label{tab:cv-posterior-estimation-results}
\centering
\begin{tabular*}{\textwidth}{@{\extracolsep\fill}lccc}
\toprule
Strategy & Log-score & MAP error & Rank score \\
\midrule
\texttt{SNPE-C}  & $-0.803 \pm 0.716$ & $0.141 \pm 0.152$ & $0.025 \pm 0.050$ \\
\texttt{FMPE}    & $-0.701 \pm 0.078$ & $0.102 \pm 0.038$ & $0.167 \pm 0.322$ \\
\texttt{NPSE}    & $2.157 \pm 0.186$ & $0.291 \pm 0.109$ & $<0.001$ \\
\texttt{DNRE} & $-1.33 \pm 0.031$ & $0.086 \pm 0.014$ & $0.0032 \pm 0.004$\\
\botrule
\end{tabular*}
\end{table*}

Given \texttt{DNRE}'s relationship to the SBI global fits of sterile neutrino models, we include \texttt{DNRE} as a posterior estimation strategy to test in this work. We also note that a better-performing posterior estimation strategy (particularly one with a better maximum \textit{a posteriori} error, see Sec.~\ref{subsec:metrics}) than \texttt{DNRE} could be used in place of Eq.~\ref{eq:dnre-logposterior} in the best fit point determination of the method described in Ref.~\cite{10.1088/2632-2153/ae040c}, further motivating its inclusion in this body of work.

\subsection{Evaluation metrics\label{subsec:metrics}}

\begin{table*}[t]
\caption{Performance of posterior estimation strategies presented in Sec.~\ref{sec:sbi-methods}, including scores for network calibration, and timing considerations for both training and MAP computation (a standardized task for each network, especially relevant for the best fit point determination and test statistic construction necessary for the method presented in Ref.~\cite{10.1088/2632-2153/ae040c}). Equivalents computed using five-fold cross validation are given in Tab.~\ref{tab:cv-posterior-estimation-results}.}\label{tab:posterior-estimation-results}
\begin{tabular*}{\textwidth}{@{\extracolsep\fill}lccccc}
\toprule%
& \multicolumn{3}{@{}c@{}}{Calibration metric} & \multicolumn{2}{@{}c@{}}{Time [s]} \\\cmidrule{2-4}\cmidrule{5-6}%
Strategy & Log-score & MAP error & Rank score & Network training & MAP computation (avg.) \\
\midrule
\texttt{SNPE-C}  & $-0.567$ & ${0.035}$ & $0.030$ & 2,864 & ${0.186}$ \\
\texttt{FMPE}  & $-0.702$ & $0.056$ & ${0.067}$ & 46,639 & $23.442$ \\
\texttt{NPSE} & ${2.160}$ & $0.302$ & $<0.001$ & 170,014 & $631.371$ \\
\texttt{DNRE} & $-1.335$ & $0.132$ & $<0.001$ & 986,375 & 503.403 \\
\midrule
\midrule
See also & Fig.~\ref{fig:log-scores} & Fig.~\ref{fig:map-errors} & Fig.~\ref{fig:rank-scores} & - & - \\
\botrule
\end{tabular*}
\end{table*}

Given a posterior density estimator $\hat{\mathbb{P}} (\theta | \mathbf{x})$, certain desirable characteristics motivate performance metrics which can be cross-compared between multiple algorithms, and should quantify the inference stratgy's precision, accuracy, and calibration in an unbiased way.

As such, for an experimental sample $\mathbf{x}_i$ simulated from oscillation parameters $\theta_i$, a good posterior estimation strategy will have a high log-score, $\log \mathbb{P}(\theta_i | \mathbf{x}_i)$, a measure of the probabilistic weight assigned at the true underlying oscillation parameters $\theta_i$. From an estimated posterior distribution, one can extract the maximum \textit{a posteriori} (MAP) $\hat{\theta}_i^\text{MAP} = \textrm{argmax}_\theta \hat{\mathbb{P}} (\theta | \mathbf{x}_i)$.\footnote{The MAP estimate also takes an important role as the best fit for SBI frequentist trials-based fitting strategies, like that presented in Ref.~\cite{10.1088/2632-2153/ae040c}} The accuracy of the MAP estimate (computed on the same grid in $U_{e4}, \Delta m_{41}^2$ as the data was generated) can be quantified by the MAP mean squared error $\Vert \theta_i - \hat{\theta}_i^\text{MAP} \Vert_2^2$ in regions where one expects the injected signal parameters to be well-reconstructed ($\log_{10} U_{e4} > -0.7$ in the case of the global fit).

\begin{figure*}[tp!]
\centering
\includegraphics[width=0.95\textwidth]{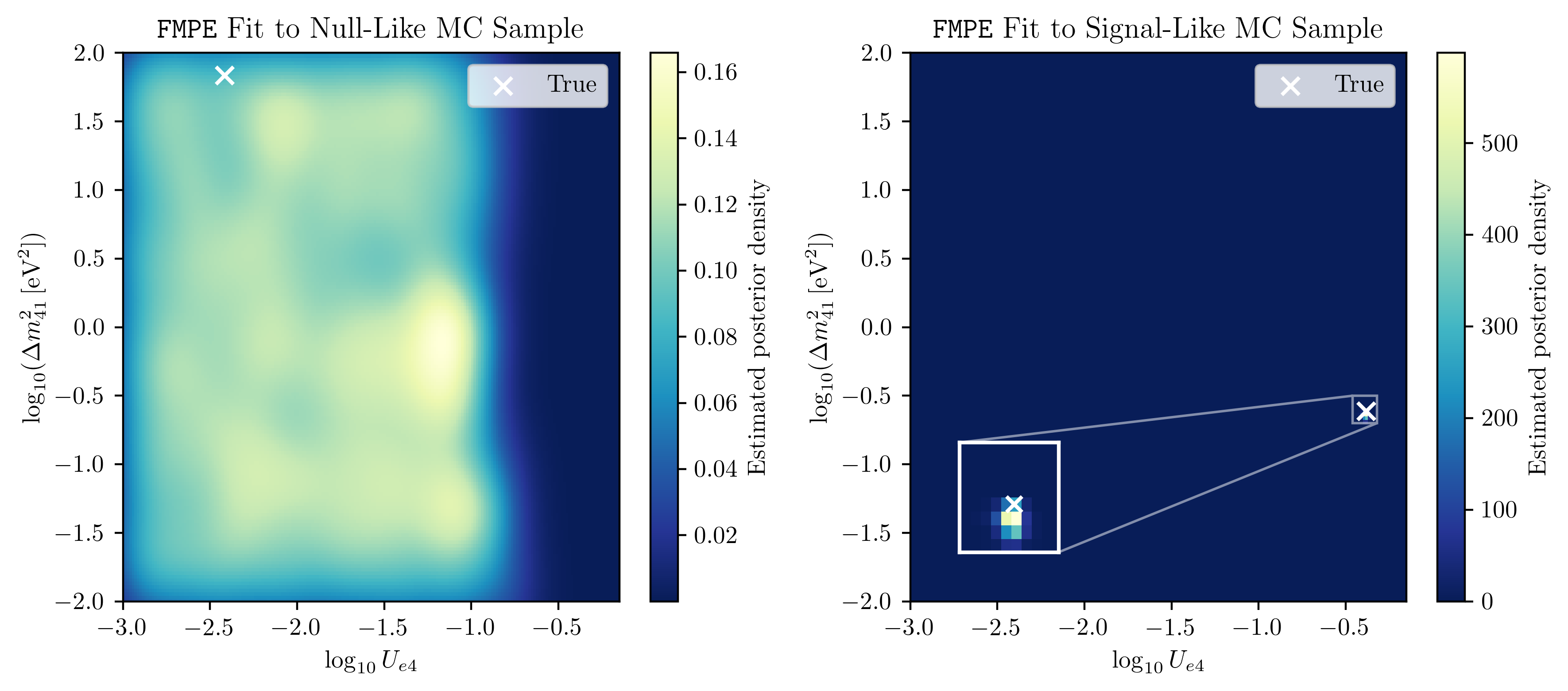}
\caption{\label{fig:mc-fits} Using simulated data, posterior density estimation performed by \texttt{FMPE} on out-of-sample null-like (left) and signal-like (right) pseudoexperimental throws. The fit to the signal sample includes an inset demonstrating excellent agreement between the sharply peaked posterior estimate and the true injected sterile oscillation parameters. These fits are to all of the electron neutrino disappearance experiments presented in Tab.~\ref{tab:experiments}. }
\end{figure*}

The calibration of the posterior estimator is determined by the rank score. Recall that in a Bayesian analysis, model parameters $\theta$ are treated as random variables, whose probability distribution is determined by the posterior. As such, given $\mathbf{x}_i$, the corresponding model parameters $\theta_i$ should be thought of as a random draw according to the posterior: $\theta_i \sim \mathbb{P} (\theta | \mathbf{x}_i)$. For the true posterior distribution, if one were to draw $N$ samples of $\theta$ from the posterior, one would expect that when these $N$ samples and the true model parameters $\theta_i$ are ranked according to their log-scores, the rank of $\theta_i$,
\begin{equation*}
    \text{rank}(\theta_i) = 1 + \sum_{j=1}^N \mathbbm{1}(\log \mathbb{P} (\theta_j | \mathbf{x}_i) > \log \mathbb{P} (\theta_i | \mathbf{x}_i) )
\end{equation*}

\noindent should be uniformly distributed. A well-calibrated posterior estimator $\hat{\mathbb{P}} (\theta | \mathbf{x}_i) $ will share this property. The measure of the rank uniformity of the true oscillation parameters is referred to as the rank score. To determine rank uniformity, we use a Kolmogorov-Smirnov test to verify that $\textrm{rank}(\theta_i) \sim \text{DUnif}(1, N+1)$. The $p$-value of the test is presented in Tab.~\ref{tab:posterior-estimation-results}.

These three metrics are the basis of comparison for the posterior density estimation strategies presented in Sec.~\ref{subsec:posterior-estimators}. Cross-validated scores across many MC realizations in the test set are summarized in Tab.~\ref{tab:posterior-estimation-results} and Fig.~\ref{fig:evaluation-summary}.

\subsection{Discussion of Flow Choice}

Based on the results of the network evaluation tasks presented in \cref{tab:cv-posterior-estimation-results,tab:posterior-estimation-results} and Fig.~\ref{fig:evaluation-summary}, the clear front-runner for a precise and accurate posterior density estimation model for the task considered in this work is \texttt{FMPE}. Across each of the five folds of the validation data, \texttt{FMPE} most consistently provides the strongest (and physically reasonable) log-scores, MAP errors, and rank scores, indicating a well-calibrated model. For the task of a standalone likelihood-free Bayesian inference of global electron-neutrino disappearance data, we recommend and proceed with \texttt{FMPE} as the chosen strategy for posterior estimation.

However, we point out that the discrete normalizing flow-based \texttt{SNPE-C} method is especially impressive in its faster training and MAP evaluation time, as well as its competitive MAP errors. As such, it is our preferred choice for upgrading the frequentist trials-based global fitting framework presented in Ref.~\cite{10.1088/2632-2153/ae040c}; namely, for use in the MAP estimation which satisfies the ranking criterion of the Feldman-Cousins test statistic (see Sec. \ref{sec:upgrading-freq}).

\section{The $\nu_e$ Disappearance Global Fit with Flow Matching Posterior Density Estimation}\label{sec:results}

\begin{figure*}[tpb!]
\centering
\includegraphics[width=0.75\linewidth]{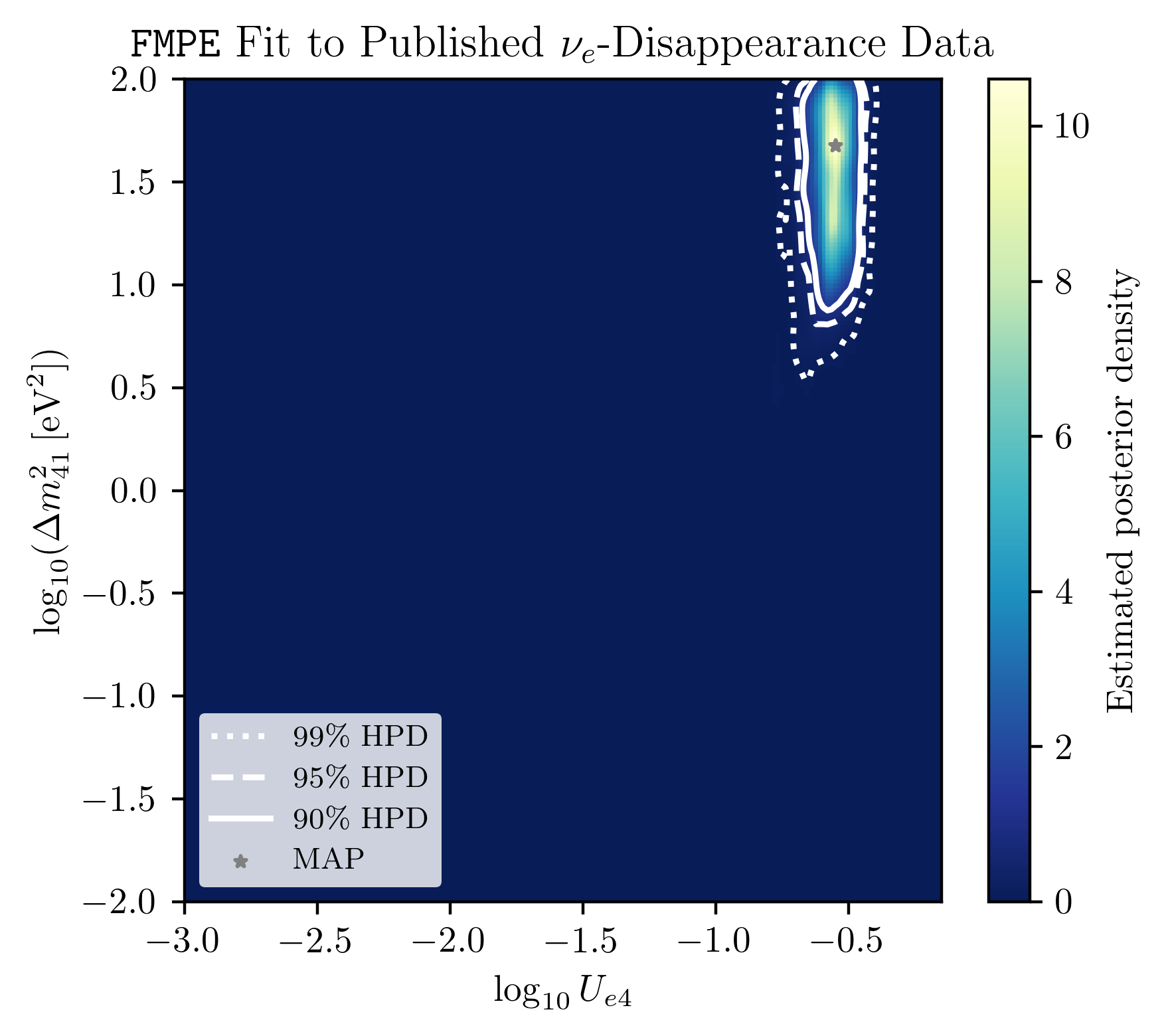}
\caption{\label{fig:true-fit-fmpe} Posterior density estimated with \texttt{FMPE} for the published $\nu_e/\overline{\nu}_e$ disappearance data in Tab.~\ref{tab:experiments}. White contours show highest-posterior-density credibility regions. The MAP estimate lies at $U_{e4}=0.28$ ($\sin^2 2\theta_{ee}=0.29$), $\Delta m_{41}^2 = 48.69\,\text{eV}^2$. Edge effects cause the normalization seen in the source experiments to taper at large $\Delta m_{41}^2$ (see discussion). This posterior approximates the MCMC result in Fig.~\ref{fig:true-fit-sblmc} and is qualitatively consistent with the frequentist allowed region in Fig.~\ref{fig:freq-fit}.}
\end{figure*}

We will now apply our methods to the electron flavor oscillation data to highlight the value of this approach.

\subsection{The Expectation for Classes of Experiments \label{subsec:precedent}}

For the classes of experiments considered in this work, previously published results provide clues as what reasonable global fits to $\nu_e$-disappearance experiments should find.  Before considering the results of our fits, let us consider the expectations.

First, consider the reactor experiments.  As discussed in Sec.~\ref{reactorfluxdiscussion},
we have renormalized the reactor data sets to unity to remove the potential confusion between a high $\Delta m^2_{41}$ signal causing rapid oscillations and theoretical uncertainty in the HM flux.  We have also followed the steps recommended by each experiment to remove sensitivity to the 5 MeV bump.
With these corrections, the reactor experiments report results that are generally consistent with the Standard Model. We point out, however, that while the newest publication from DANSS sees a $2.3\sigma$ preference for 3+1, with a best fit point at $(\sin^2 2 \theta_{ee}, \Delta m_{41}^2) = (0.07, 0.34\,\text{eV}^2)$ \cite{DANSS}, the dataset implemented in \texttt{sblmc} is significantly smaller and consistent with the Standard Model \cite{Skrobova2023}.  The RENO+NEOS search excludes even the best fit point of the larger set at $95\%$ confidence \cite{PhysRevD.105.L111101}.  Therefore, taken as a whole, we expect our treatment of the reactor experiments to overall exclude the 3+1 model over the parameter space where they are sensitive.

One the other hand, the source experiments are expected to collectively prefer 3+1 over the Standard Model, with null rejection at $\gtrsim 5\sigma$ significance when taken together \cite{cadeddu2025reassessinggalliumanomalyusing}. This is a strikingly large signal.  No SM source for the deficit seen in these experiments has been identified.  The $\nu_e$ fluxes from monoenergetic electron-capture-decaying isotopes like argon-37 and chromium-51 are well-modeled, and counting of germanium-71 atoms in liquid gallium is reliable. 

Recall that the 3+1 model we are testing requires the same oscillation parameters for $\nu_e$ and $\bar \nu_e$ fluxes.   Thus, given this disagreement in source versus reactor results, it is interesting to look to the accelerator-based data set, which is also from a $\nu_e$ flux.   While those results reported are not statistically very strong, they are consistent with null.

Overall, the source experiments are expected to dominate the global fit, but the significance will be reduced  due to exclusions from the reactor and accelerator experiments.

Given these expectations, which were guided by the published frequentist results, let us proceed to consider what the SBI-based Bayesian global fit reports based on publicly available experimental data presented the appendix (in Fig.~\ref{fig:published}).

\subsection{The Simulation-Based Inference Global Fit for Simulated and Published Results}

Sample fits using \texttt{FMPE} on Monte Carlo pseudodata are given by Fig.~\ref{fig:mc-fits}. The fit of the trained \texttt{FMPE} posterior density estimator to published $\nu_e$ and $\bar \nu_e$ disappearance data is given by Fig.~\ref{fig:true-fit-fmpe}.   One observes that in the global fit, the signal from the source experiments more than off-sets the exclusion from the reactor experiments leaving an allowed region.   We discuss how each class of experiment contributed below.

\subsection{Fits to Each Experimental Class}

\begin{figure}[tp!]
\centering
\includegraphics[width=0.95\columnwidth]{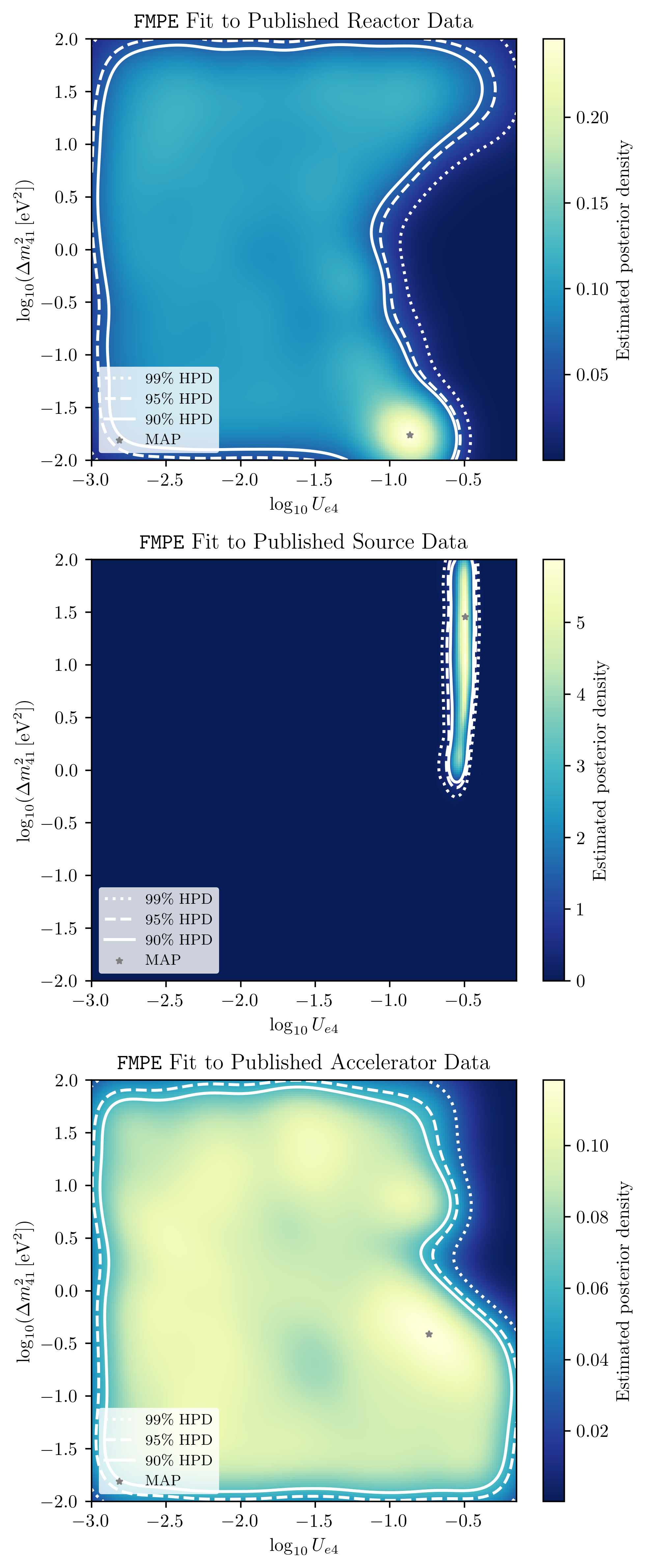}
\caption{\label{fig:fmpe-split-fits}Fits to published data from each of the three experiment classes (reactors, including STEREO, PROSPECT, DANSS, and NEOS; sources, including SAGE/GALLEX and BEST; accelerators, including the Karmen-LSND cross section analysis) using \texttt{FMPE}. These agree qualitatively with the frequentist allowed regions in Fig.~\ref{fig:diff-exps-freq}.}
\end{figure}

\begin{figure}[tp!]
    \centering
    \includegraphics[width=0.95\linewidth]{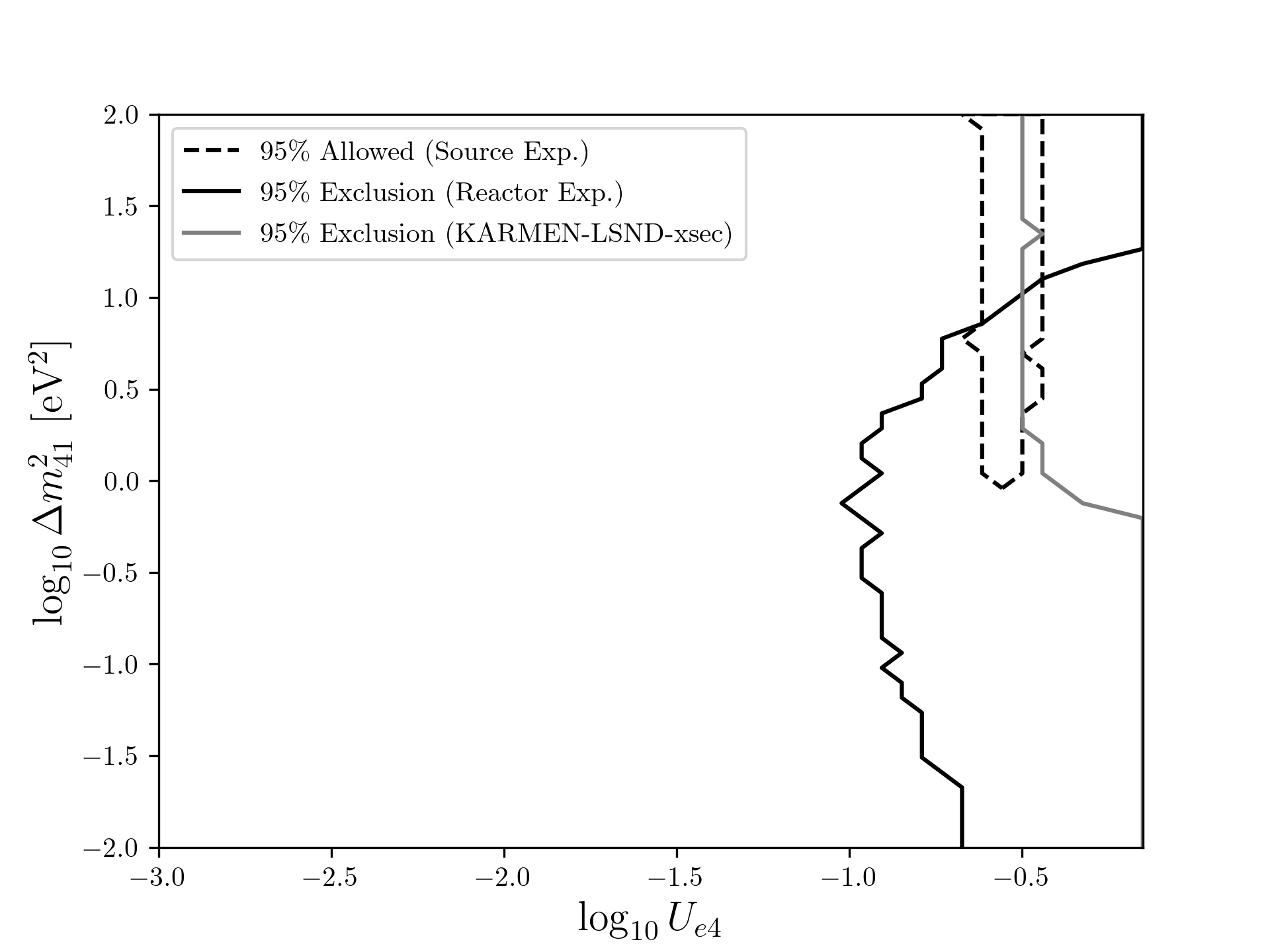}
    \caption{Acceptance regions for each experiment class within the frequentist framework, computed using the DNRE posterior evaluation.}
    \label{fig:diff-exps-freq}
\end{figure}

Fits to each of the experiment classes (reactor, source and accelerator) are shown in Fig.~\ref{fig:fmpe-split-fits}. Based on the overview presented in Sec.~\ref{subsec:precedent}, likelihood-free inference of the three classes are in good agreement with historical precedent. In particular, fits to the reactors and the accelerators show no significant preference for a 3+1 signal, while the source experiments show a strong preference for a sterile neutrino. Moreover, the allowed region preferred by the source experiments remains in the global picture (Fig.~\ref{fig:true-fit-fmpe}), though the permissible model parameter space is limited to $\Delta m_{41}^2 \gtrapprox 10\,\text{eV}^2$, driven by the exclusion set by the reactor experiments. Moreover, the Bayesian fits are in good agreement with their frequentist counterparts, using the method of Ref.~\cite{10.1088/2632-2153/ae040c}; see Fig.~\ref{fig:diff-exps-freq}. 

\section{Upgrading the Trials-Based Frequentist Global Fit}\label{sec:upgrading-freq}
In a frequentist global fit, the most powerful test-statistic is the likelihood ratio $\hat r(x, \theta, \hat\theta) = \frac{\mathbb{P}(x|\theta)}{\mathbb{P}(x|\hat\theta)}$, where $\hat\theta$ denotes the parameter values that maximize the likelihood. In our frequentist muon neutrino disappearance analysis \cite{10.1088/2632-2153/ae040c}, finding the maximum likelihood estimator (MLE) is challenging within the SBI pipeline, and we instead take $\hat\theta$ to maximize the posterior distribution which we obtain from DNRE. Eq. \ref{eq:dnre-logposterior} provides a prescription for posterior estimation from DNRE likelihood ratios. This is crucially dependent on hyper-parameter $M$, which is the number of samples drawn from the prior over which likelihood ratios are averaged. Specifically, \begin{equation}
    \frac{\mathbb{P}(x|\theta)}{\mathbb{P}(x)} = \frac{1}{M}\sum_i^M \hat r(x|\theta, \theta_i').
\end{equation}
Increasing $M$ improves coverage accuracy \cite{10.1609/aaai.v38i18.30018}, but comes at the cost of significant reduction in computational efficiency. 
To balance these tradeoffs, in our frequentist study we fixed $M=600$. With this choice, evaluating likelihood ratios over the grid of parameter space for a single experiment sample took $\mathcal{O}(10)$s. 
\par While this is a significant improvement from computing the test-statistic using traditional Feldman-Cousins, applying a normalizing flow yields an additional speed-up (Table \ref{tab:posterior-estimation-results}). Although the network training time for the normalizing flows is notably greater than for DNRE, the largest bottleneck in the frequentist study is MAP generation. The strategy with the fastest MAP computation average is \texttt{SNPE-C} by a factor of around 200. Given the higher log and rank score and lower MAP error (Table \ref{tab:posterior-estimation-results}), we anticipate substituting \texttt{SNPE-C} for the DNRE posterior estimation into the frequentist pipeline will yield both more accurate MAP estimates and greater computational efficiency. 
\par
\begin{figure}[tp!]
\centering
\includegraphics[width=0.95\columnwidth]{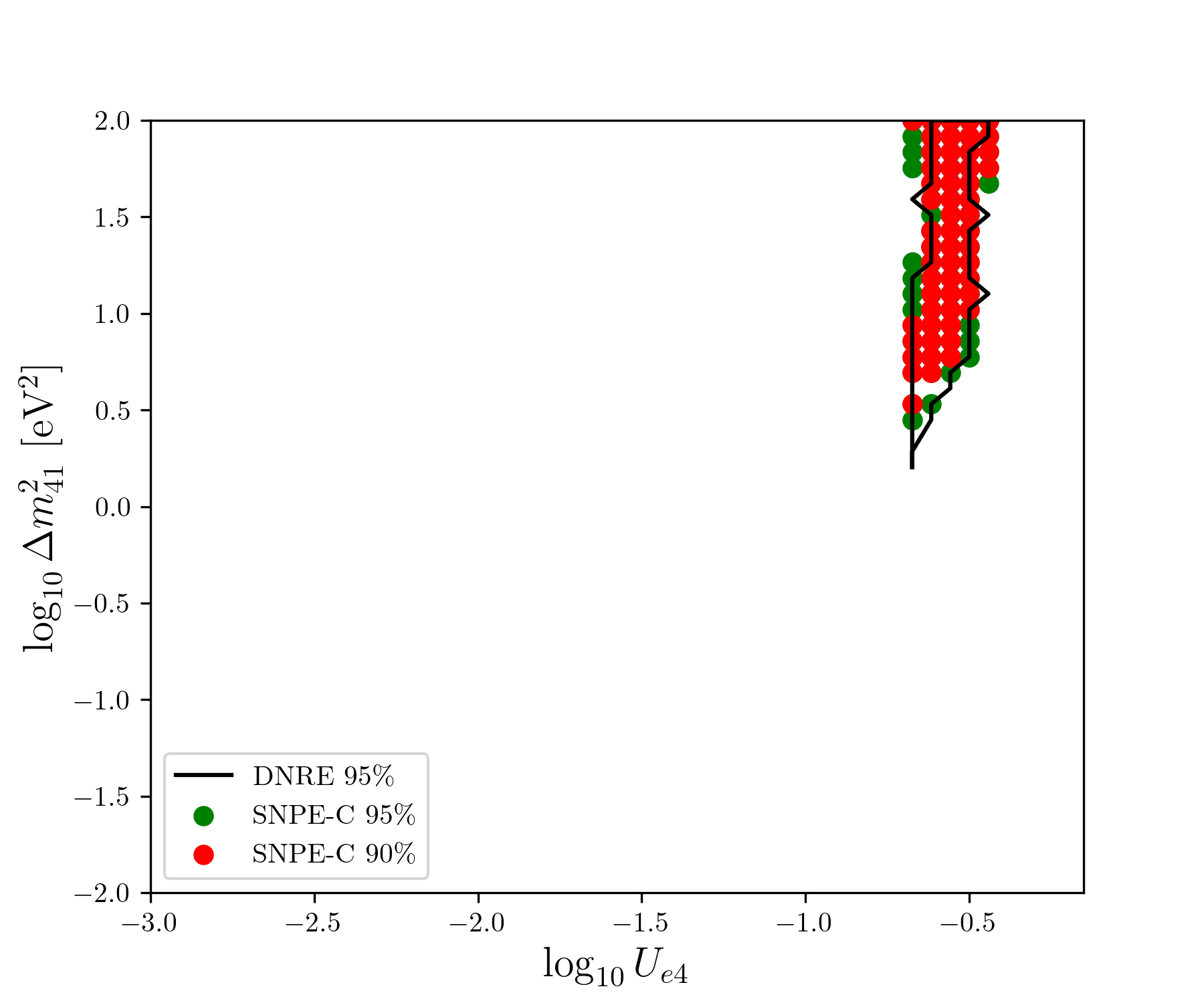}
\caption{\label{fig:freq-fit} Frequentist allowed regions for the global $\nu_e/\overline{\nu}_e$ disappearance fit, using the \texttt{SNPE-C} posterior density to estimate the MAP. Red and green denote the 90\% and 95\% confidence regions. The \texttt{DNRE}-derived 95\% confidence level is also shown, following the procedure in \cite{10.1088/2632-2153/ae040c}. These regions can be qualitatively compared with the Bayesian credibility regions from \texttt{FMPE} (Fig.~\ref{fig:true-fit-fmpe}) and the high fidelity MCMC global fit in Fig.~\ref{fig:true-fit-sblmc}.
}
\end{figure}

 To find the MAP using \texttt{SNPE-C}, we train the flow on the same data used to train the DNRE, which is still needed for likelihood ratio estimation. For each pseudo-experiment, we sample 50,000 points from the posterior. The MAP is the parameter point with the highest log-probability. The frequentist confidence levels using \texttt{SNPE-C} and DNRE to compute the MAP are shown in Fig. \ref{fig:freq-fit}. Both methods yield roughly the same allowed region. The \texttt{SNPE-C} method finds a MAP located at $U_{e4} = 0.31$, $\Delta m_{41}^2 = 13.36$eV$^2$, while DNRE finds a MAP located at $U_{e4} = 0.28$, $\Delta m_{41}^2 = 18.42$eV$^2$.

\par On the same grid and for the same data sample, evaluating likelihood ratios over the parameter space using \texttt{SNPE-C} for posterior estimation takes $\sim 0.1$s, corresponding to a $\mathcal{O}(100)$-fold speed up. The use of \texttt{SNPE-C} also removes the dependence on the hyper-parameter $M$, which no longer needs to be optimized. 
\par A full 3+1 neutrino global fit involves scanning over three independent parameters, greatly expanding the parameter space. As it stands, performing a three-parameter fit using DNRE-based posterior estimation is computationally infeasible. The adoption of \texttt{SNPE-C} overcomes this limitation, making such analyses possible. 

\begin{figure}
\centering
\includegraphics[width=0.48\textwidth]{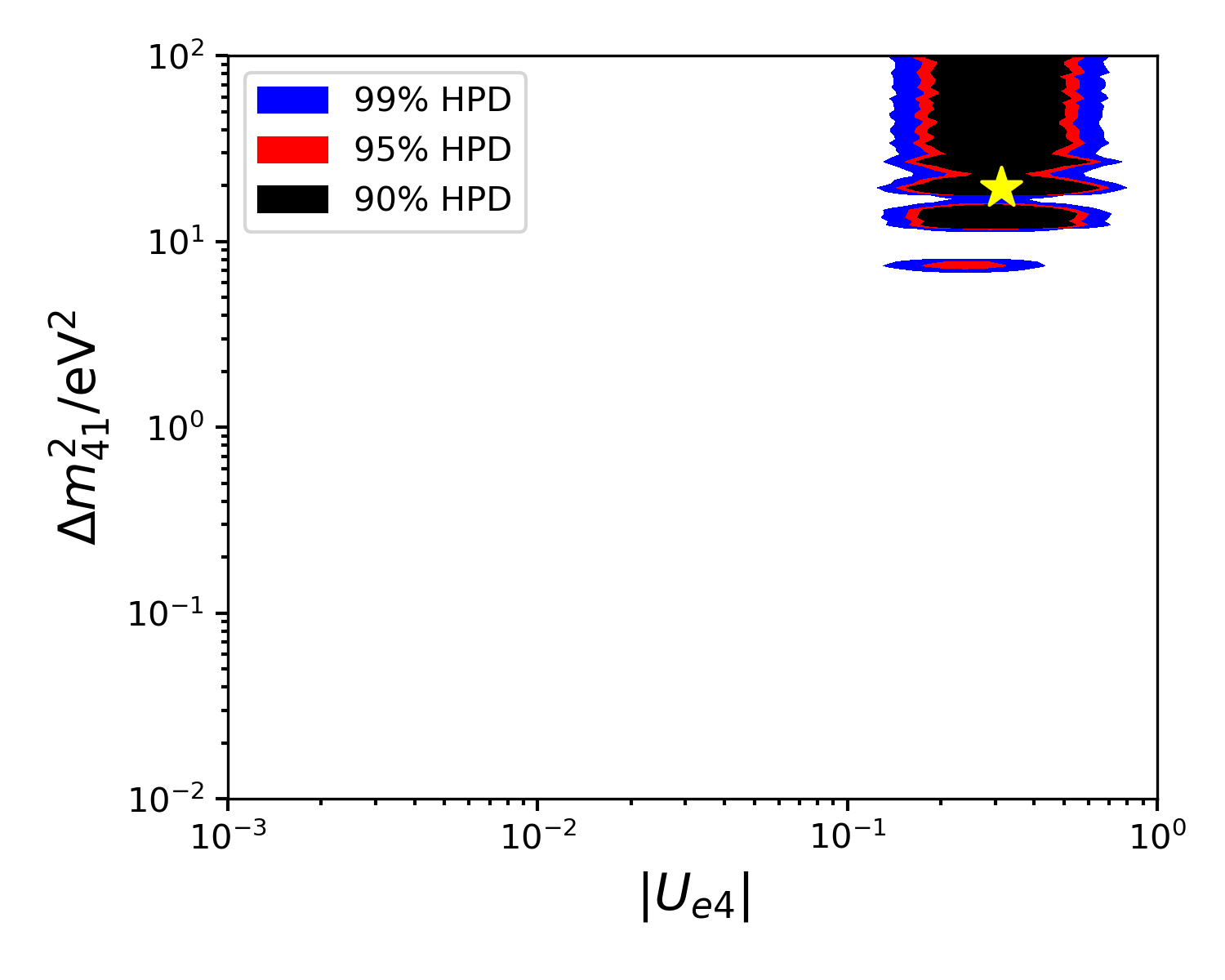}
\caption{\label{fig:true-fit-sblmc}
Posterior density estimated with \texttt{sblmc} for the published $\nu_e/\overline{\nu}_e$ disappearance data in Tab.~\ref{tab:experiments}. Colored regions indicate highest-posterior-density credibility intervals. The MAP estimate occurs at $U_{e4}=0.31$ ($\sin^2 2\theta_{ee}=0.35$), $\Delta m_{41}^2 = 13.7\,\text{eV}^2$. See Ref.~\cite{Diaz:2019fwt} for details of the method. This is the exact posterior approximated in Fig.~\ref{fig:true-fit-fmpe} and is qualitatively consistent with the features in Fig.~\ref{fig:freq-fit}.}

\end{figure}

\section{Discussion}\label{sec:discussion}

\subsection{Fit Comparisons}

In the case of the global fit to all available electron-flavor disappearance data, excellent topological agreement can be seen between the directly comparable true Markov Chain Monte Carlo fit (Fig.~\ref{fig:true-fit-sblmc}) and \texttt{FMPE} (Fig.~\ref{fig:true-fit-fmpe}). Moreover, amortized inference with flow matching is significantly faster ($\sim 10\,\text{s}$) than the posterior sampling performed by \texttt{sblmc} ($\sim 10^4\,\text{s}$ for convergence). The identified maximum \textit{a posteriori} from the high-fidelity Bayesian fit is in good agreement with that of the flow matching, though we note a slight discrepancy in the found value of $\Delta m_{41}^2$. This is unsurprising -- in frequentist fits, the shallow minimum of the $\Delta \chi^2$ landscape due to the influence of the gallium anomaly on the fit is a known artifact. Poor detector segmentation of the source experiments means poor sensitivity to the mass splitting. Topological agreement can also be observed between the SBI-based global estimated posterior in Fig.~\ref{fig:true-fit-fmpe} and the SBI-powered trials-based method (developed in Ref.~\cite{10.1088/2632-2153/ae040c}) in Fig.~\ref{fig:freq-fit}. There is almost exact agreement between the MAP found by the true Markov Chain Monte Carlo fit and that from \texttt{SNPE-C} (Sec. \ref{sec:upgrading-freq}). With one method a Bayesian approach and the other a freuqentist approach, the fit results are not directly comparable but do indicate similar localized signal preference driven by the source experiments.

Excellent topological agreement in preferred model parameter space for experiments split by class (reactors, sources, and accelerators) can be seen between the Bayesian (Fig.~\ref{fig:fmpe-split-fits}) and frequentist (Fig.~\ref{fig:diff-exps-freq}) SBI approaches. Both the reactor experiments and accelerator experiments agree in their regions of exclusion, and the source experiments agree in their preference for a normalization effect consistent with the gallium anomaly. Comparisons of both SBI approaches for single experiments are also shown in Fig.~\ref{fig:single-experiment-fits}. While STEREO, PROSPECT, DANSS, and BEST appear to be in good agreement with one another, the frequentist fit for NEOS draws a closed 95\% CL, while the Bayesian fit seems to prefer a result more consistent with null. Regions of high estimated posterior probability assigned by \texttt{FMPE} seem to agree with the region preferred by the frequentist fit. It is worth noting that at a 99\% confidence level, the NEOS frequentist fit draws an exclusion in agreement with the Bayesian counterpart. A similar story emerges in the comparisons of SAGE and GALLEX; despite the weak exclusions drawn by \texttt{FMPE}, regions of parameter space preferred by the fit are in good agreement with the confidence level drawn by the method from Ref.~\cite{10.1088/2632-2153/ae040c}.

\subsection{Limitations}

Given that field expectations are usually presented in a frequentist framework, care must be taken when conducting Bayesian analyses of particle physics data. Credibility regions are not subject to the same coverage guarantees as frequentist confidence levels \cite{4d625c73-e285-3ead-ae1f-415e8ad60311, Talts:2018zdk}. The choice of prior (defined implicitly by the model parameter space covered by the training data) can also strongly bias the posterior, credibility regions, and maximum \textit{a posteriori} in low-statistics or degenerate regions. Hard parameter boundaries (introduced via the uniform prior in this study) can give rise to edge effects that a density estimator may amplify. In higher parameter space dimensions, posterior mass can accumulate on thin, highly non-intuitive manifolds due to the concentration-of-measure phenomenon; low-dimensional marginal plots can therefore be misleading, hiding multimodality, degeneracy ridges, and the fact that MAP estimates are often unrepresentative of typical posterior mass \cite{Giannopoulos2000-cu}. Communicating the geometry of the posterior thus becomes inherently challenging.

While conducting inference with flows is generally faster than a high-fidelity MCMC framework, significant up-front costs in data generation and network training must be considered. Before any analysis can be performed, a sufficient amount of training data must be simulated, inefficient for one-off analyses. There is also an amortization mismatch: once trained, the network is fixed, and changes to the prior, experimental configuration, or simulator require retraining of the flow. As demonstrated in this study, posterior estimation quality depends strongly on network architecture. 

Finally, there are challenges of parameter inference using simulation-based inference generally. Biases or mismodeling in the simulator propagates errors into the posterior estimator; anecdotally, however, we point out that unexpected behaviors of the flows' helped us to narrow down previously undiscovered bugs in the simulation. When observed data is fed to a flow, and that data may fall outside the training distribution by way of, for instance, unmodeled systematic effects, possibly overconfident (or even nonsensical) posteriors can arise \cite{10.1088/2632-2153/ae040c}. Inference failures, especially in tails or low-likelihood regions, can be subtle and difficult to diagnose, requiring specialized tests. 

In spite of these limitations, amortized posterior density estimation still remains a powerful tool in the analysis pipeline. When carefully validated and used alongside frequentist methods, they can illuminate complimentary structure in the data and provide rapid exploratory insights impractical with sampling-based methods alone. Methods presented in this work are not aimed to replace existing high-fidelity techniques; instead, they are additional tool in the arsenal of global fitters to serve as an additionoal diagnostic and interpretive tool, help to identify modeling issues, and rapidly iterate on complex particle physics models. 

\section{Conclusion}\label{sec:conclusion}

In this work, we performed a comprehensive evaluation of modern neural network-based posterior estimation techniques for sterile neutrino oscillation analyses via electron-flavor disappearance, comparing sequential neural posterior estimation, flow matching, and diffusion-based neural posterior score estimation against a direct amortized neural likelihood-ratio baseline. Balancing considerations to accuracy, precision, and computational efficiency, we recommend flow matching for the purposes of sterile neutrino global fits, and recommend sequential neural posterior estimation to improve a previously developed frequentist global fit pipeline. Together, techniques considered in this work enable rapid and flexible global fits retaining compatibility with established methods.

Using these posterior estimators, we performed a global analysis of electron-flavor disappearance data collected from reactor, source, and accelerator experiments, representative of the complexities of a full-fledged 3+1 global fit. By performing fits stratified by experiment and experiment class, we illustrate how neural posterior estimators (with an amortized data generation cost) can help to disentangle the relative contributions of each dataset revealing how the global 3+1 picture may fit together. Importantly, we showed that the resulting posterior surfaces closely match both high-fidelity Bayesian MCMC sampling and frequentist simulation-based inference approaches. This agreement provides evidence that neural networks can conduct both accurate and interpretable inferences at a fraction of the computational cost.

Taken together, our results indicate that flow-based posterior estimation offers a promising and practical complement to existing analysis paradigms in neutrino oscillation physics, groundwork for an end-to-end simulation-based inference 3+1 global fit, and clues for analysis methods of more complex beyond Standard Model scenarios. While care must be taken in prior specification, calibration, and validation, these methods provide a powerful means of accelerating global fits, enriching the interplay between Bayesian and frequentist perspectives. As sterile neutrino searches continue to evolve and datasets grow in complexity, neural posterior estimation offers a valuable path toward more flexible, scalable, and transparent inference pipelines capable of supporting the next generation of precision oscillation analyses.

\bmhead{Acknowledgements}
We are grateful to Phil Harris at MIT and Gabriel Collin at the University of Adelaide for helpful discussions at the birth of this project. JV, JW, JH, and JC thank MIT for support on this project. This material is based upon work supported in part by the National Science Foundation Graduate Research Fellowship under Grant No. 2141064.
\appendix
\section{Frequentist Acceptance Regions by Experiment Class}
Allowed and exclusion regions for each class of experiment at $95\%$ confidence, shown in  Fig. \ref{fig:diff-exps-freq}. 

\section{Published Data Distributions}\label{app:datadists}

Published data is made available in Fig.~\ref{fig:published} for convenience.

\begin{figure*}[t]
\centering

% ======= Row 1 =======
\begin{subfigure}[b]{0.48\textwidth}
    \includegraphics[width=\linewidth]{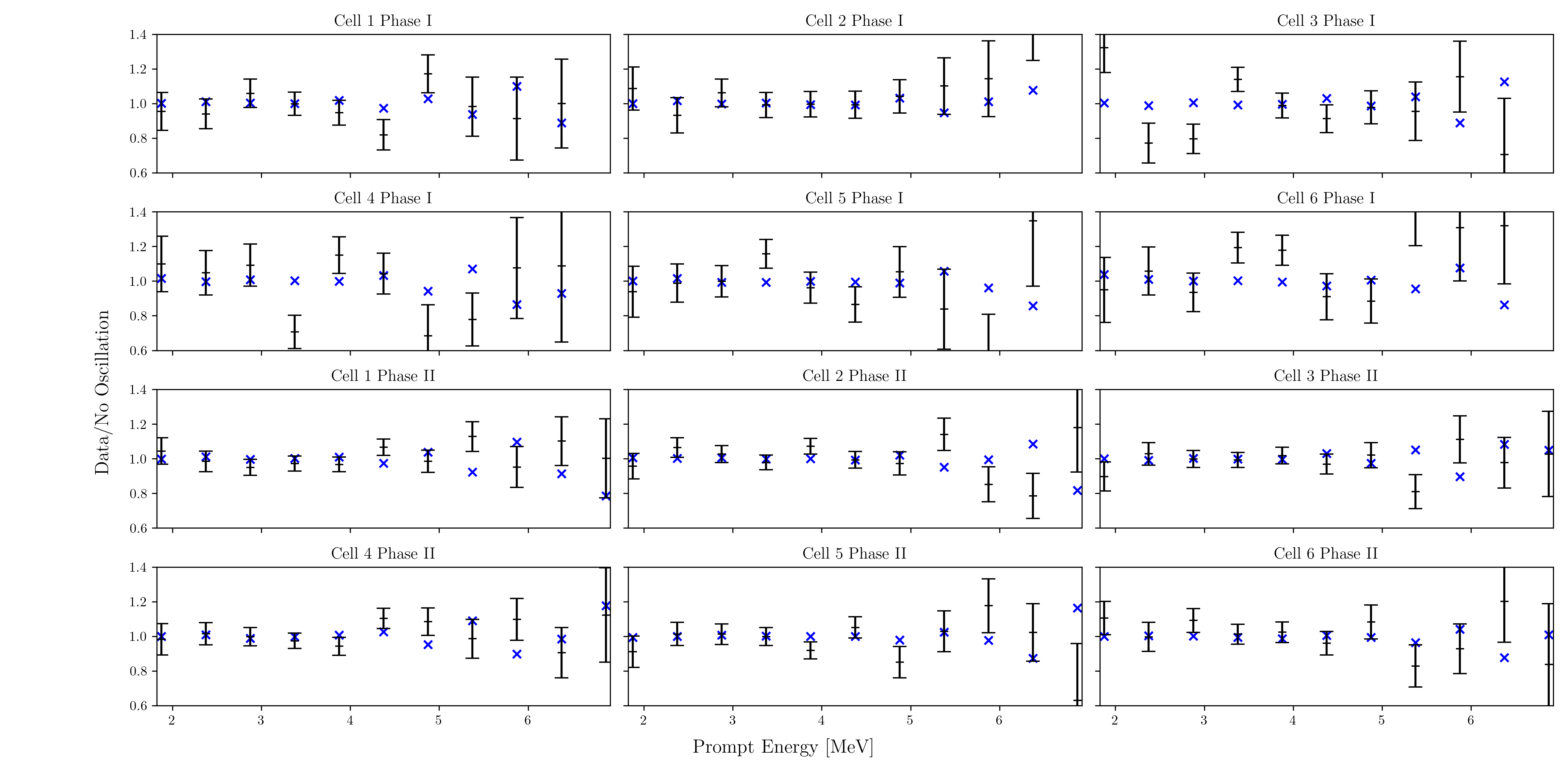}
    \caption{Ratio of data to the no-oscillation model for STEREO Phase I and II, distributed among 6 cells \cite{stereocollaboration2020antineutrino}.}
    \label{fig:stereo_data_dists}
\end{subfigure}
\hfill
\begin{subfigure}[b]{0.48\textwidth}
    \includegraphics[width=\linewidth]{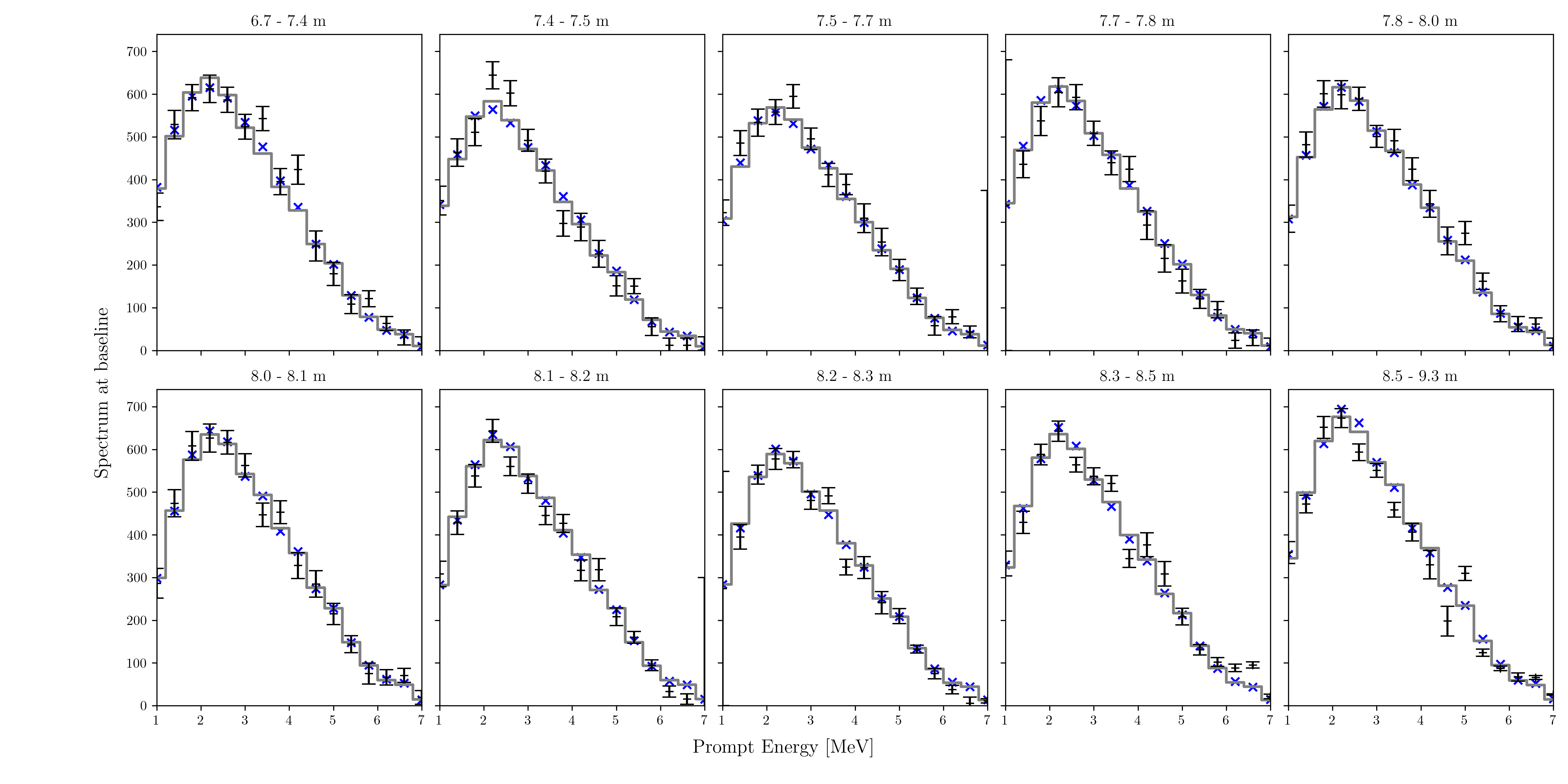}
    \caption{PROSPECT predicted spectrum at 10 different baselines, weighted by the ratio of total observed and predicted event rate in each energy bin \cite{prospectResults}.}
    \label{fig:prospect_data_dists}
\end{subfigure}

% ======= Row 2 =======
\begin{subfigure}[b]{0.48\textwidth}
    \includegraphics[width=\linewidth]{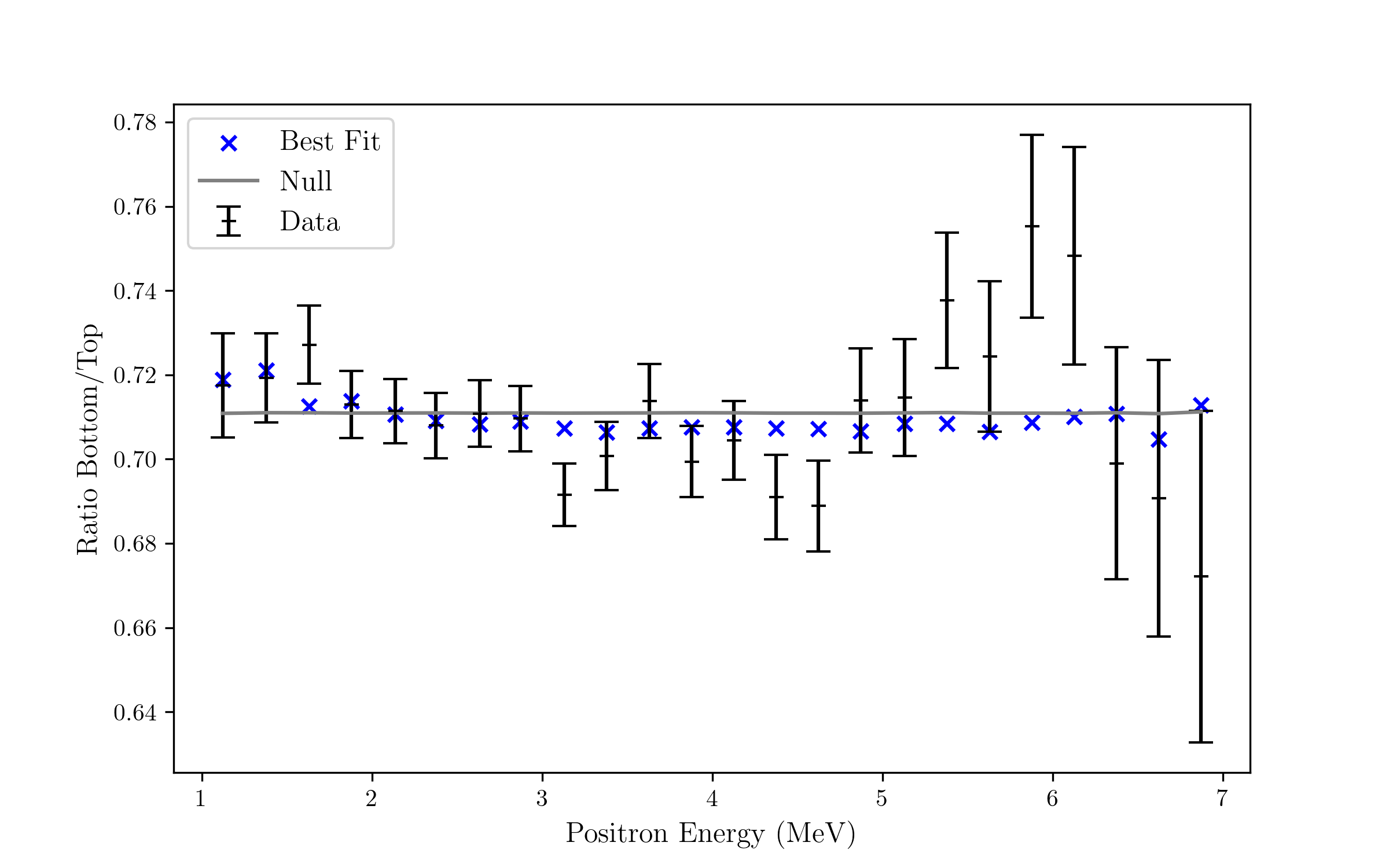}
    \caption{Ratio of positron energy at the bottom and top detector positions in DANSS \cite{DANSS}.}
    \label{fig:danss_data_dists}
\end{subfigure}
\hfill
\begin{subfigure}[b]{0.48\textwidth}
    \includegraphics[width=\linewidth]{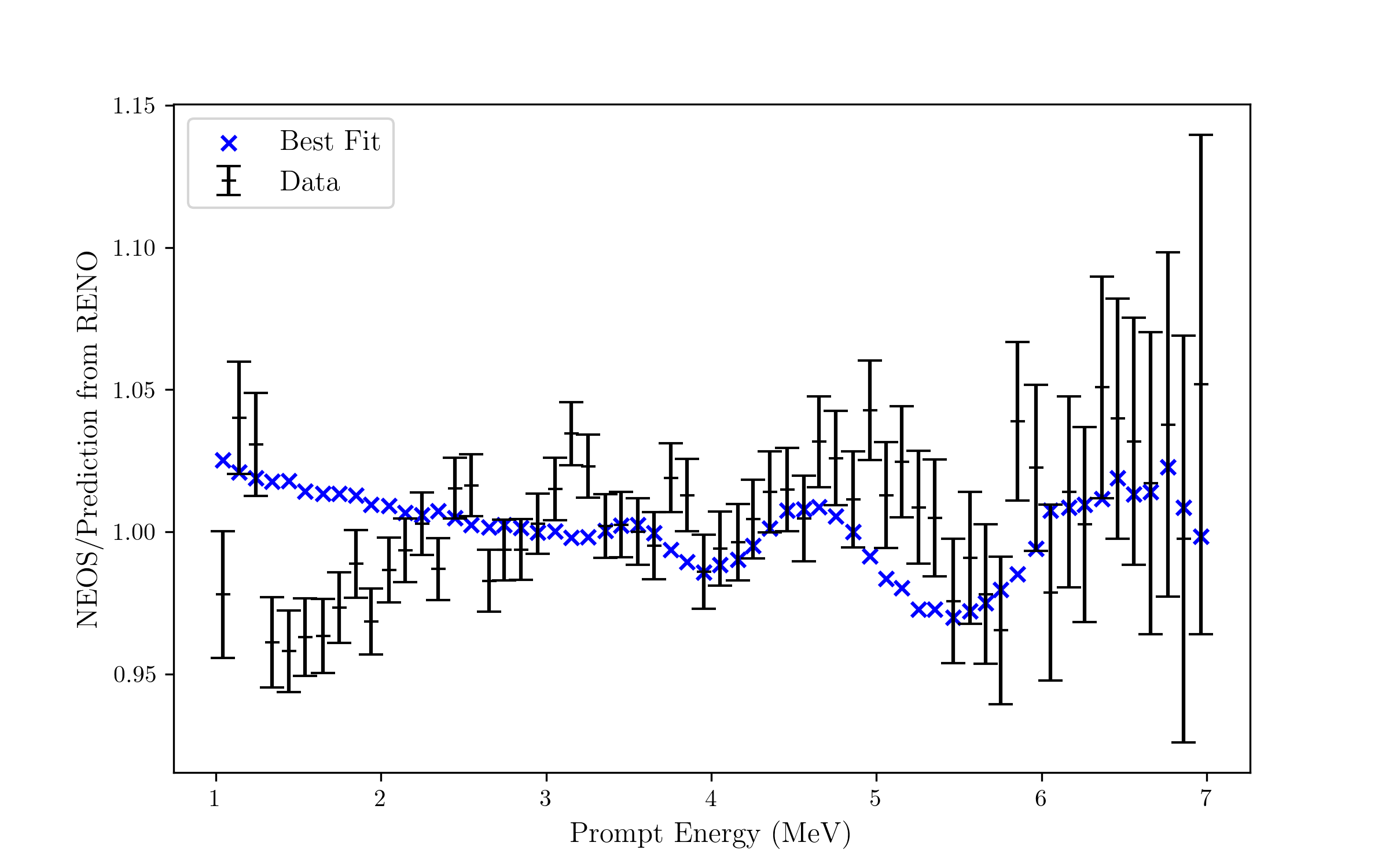}
    \caption{Ratio of NEOS and predicted RENO prompt energy spectra. The RENO $\bar{\nu}_e$ flux is used to minimize NEOS dependence on reactor flux \cite{PhysRevD.105.L111101}.}
    \label{fig:neos_data_dists}
\end{subfigure}

\caption{%
Published data (black) and averaged \texttt{sblmc} realizations generated at the published best fit (blue) for each experiment. 
For some experiments, the prediction under the null hypothesis (no sterile oscillation) is shown in gray.
}
\label{fig:published}
\end{figure*}

\begin{figure*}[t]
\ContinuedFloat
\centering

% ======= Row 3 =======
\iffalse
\begin{subfigure}[b]{0.48\textwidth}
\centering
\resizebox{\linewidth}{!}{
\begin{tabular}{c c c c c}
    \hline
    & GALLEX-1 & GALLEX-2 & SAGE-1 & SAGE-2 \\
    \hline
    Data & 1.0  & 0.81 & 0.95 & 0.79 \\
    Errors & 0.1 & 0.1 & 0.12 & 0.1 \\
    Best Fit & 0.78 & 0.77 & 0.77 & 0.76 \\
    \hline
\end{tabular}}
\caption{Ratio of measured to expected $^{71}$Ge event rates in the Gallium experiments \cite{SAGE:2009eeu}. Data presented here corresponds to a reanalysis of the Gallium data performed with updated cross section calculations \cite{PhysRevC.80.015807}.}
\label{fig:gallium_data_dists}
\end{subfigure}
\fi
\begin{subfigure}[b]{0.48\textwidth}
    \includegraphics[width=\linewidth]{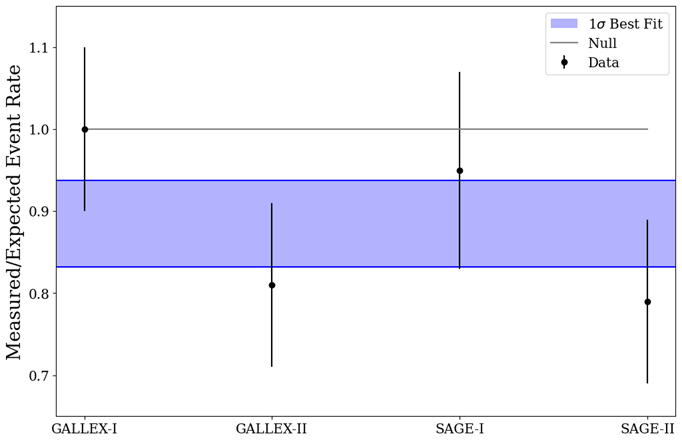}
    \caption{Ratio of measured to expected $^{71}$Ge event rates in the Gallium experiments \cite{PhysRevC.73.045805}.}
    \label{fig:gallium_data_dists}
\end{subfigure}
\hfill
\begin{subfigure}[b]{0.48\textwidth}
    \includegraphics[width=\linewidth]{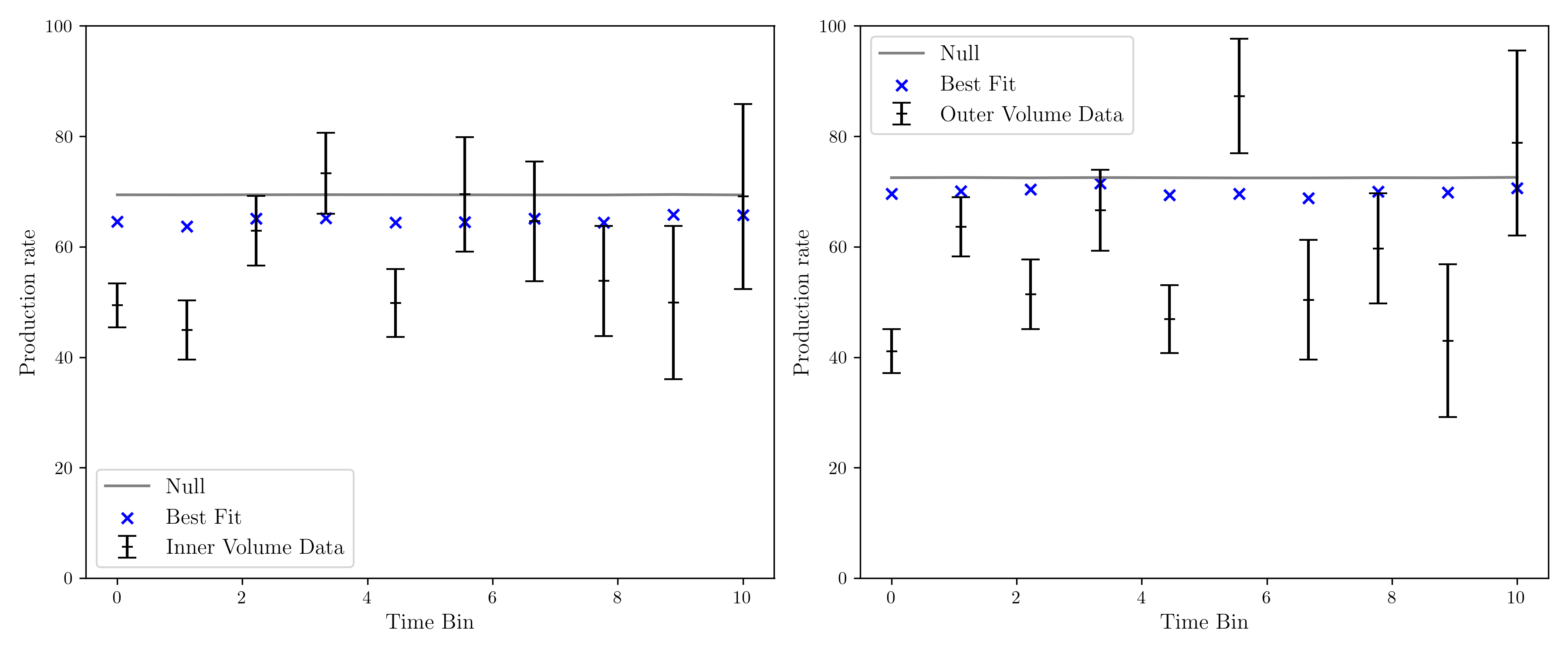}
    \caption{Production rate in $^{71}$Ge atoms/day normalized by the reference time for the inner (left) and outer (right) volumes in BEST \cite{PhysRevLett.128.232501}.}
    \label{fig:best_data_dists}
\end{subfigure}

% ======= Row 4 =======
\begin{subfigure}[b]{0.48\textwidth}
    \includegraphics[width=\linewidth]{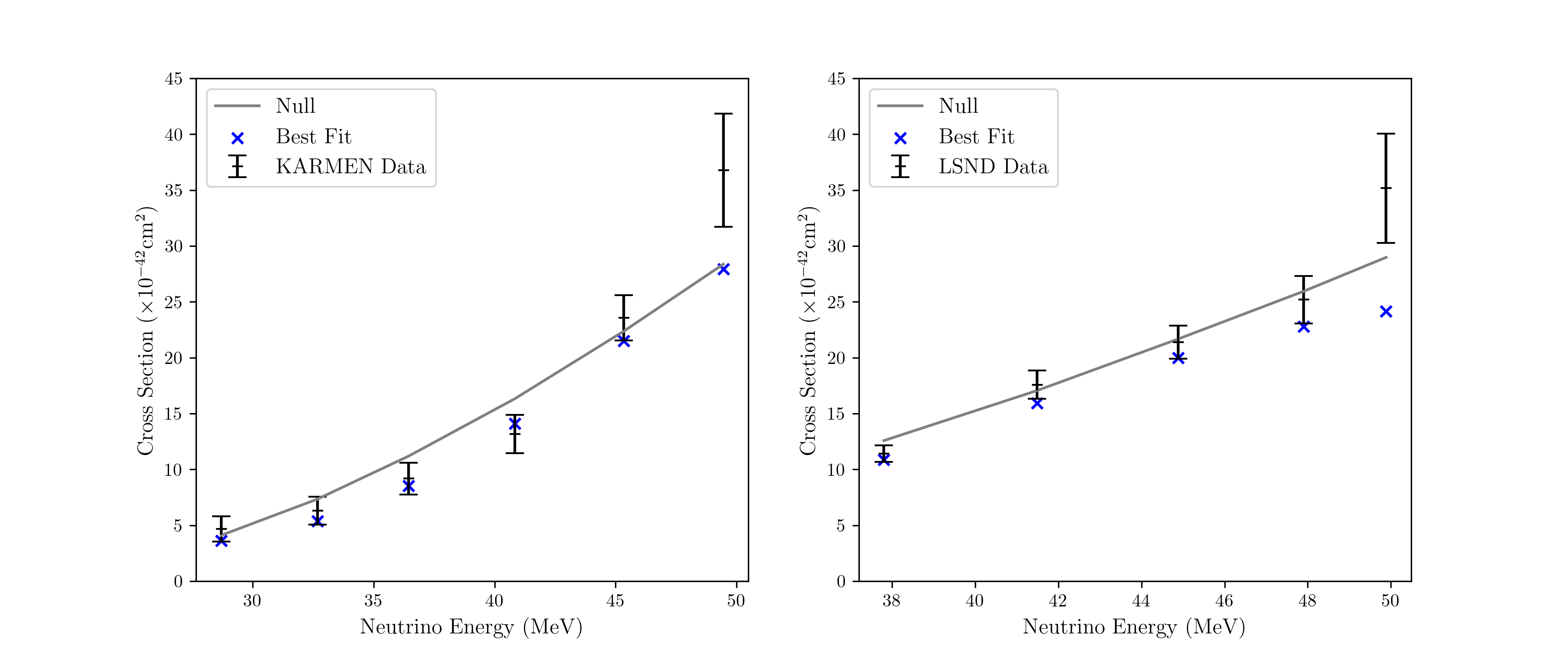}
    \caption{KARMEN (left) and LSND (right) cross section measurements at various neutrino energies \cite{Conrad:2011ce}.}
    \label{fig:klx_data_dists}
\end{subfigure}

\caption{(continued)}
\end{figure*}

\section{Single-Experiment Fits\label{sec:single-experiment-fits}}

\begin{figure*}[tp!]
\centering
\begin{subfigure}[b]{0.32\textwidth}
    \includegraphics[width=\linewidth]{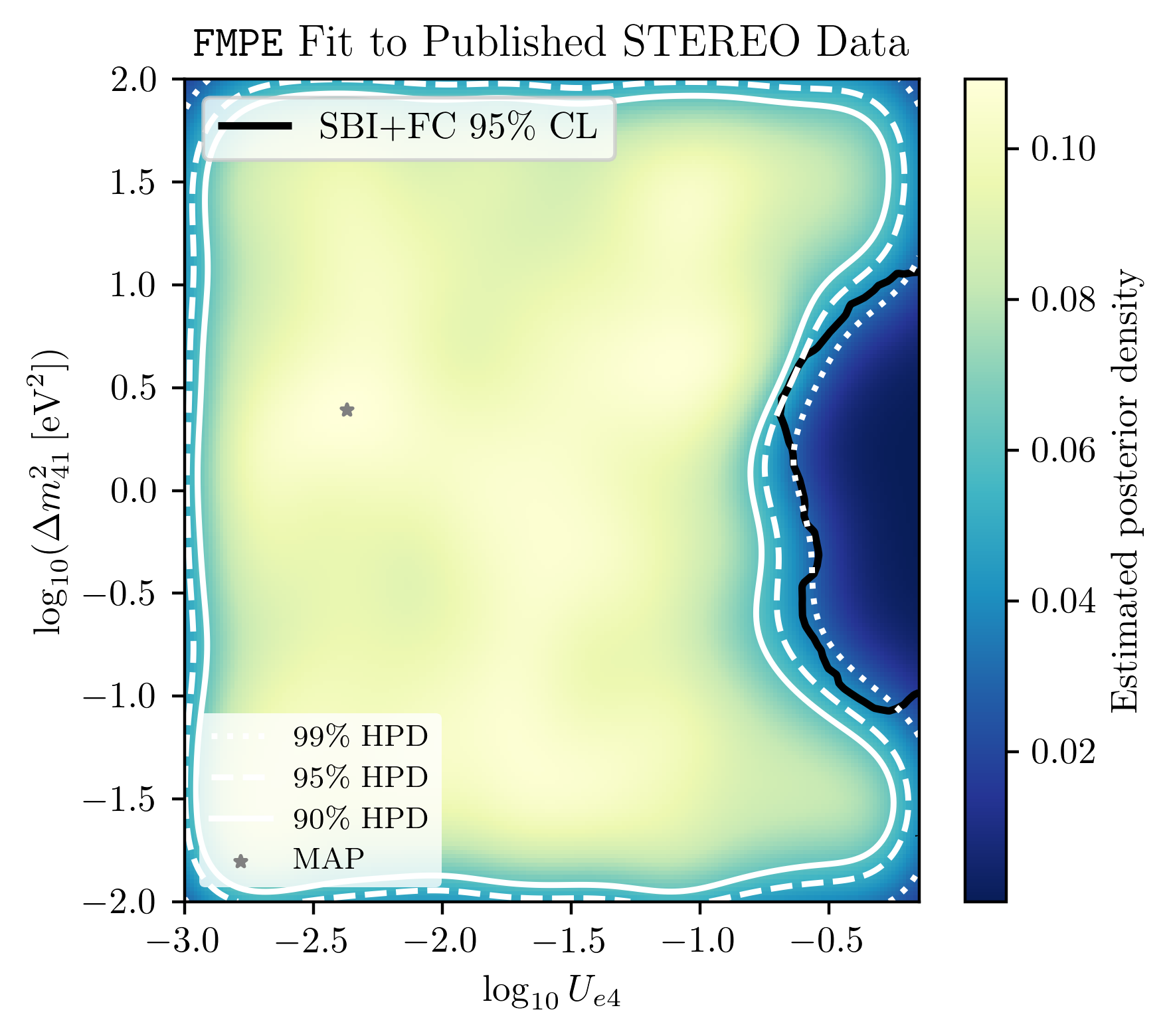}
    \caption{STEREO}
\end{subfigure}
\begin{subfigure}[b]{0.32\textwidth}
    \includegraphics[width=\linewidth]{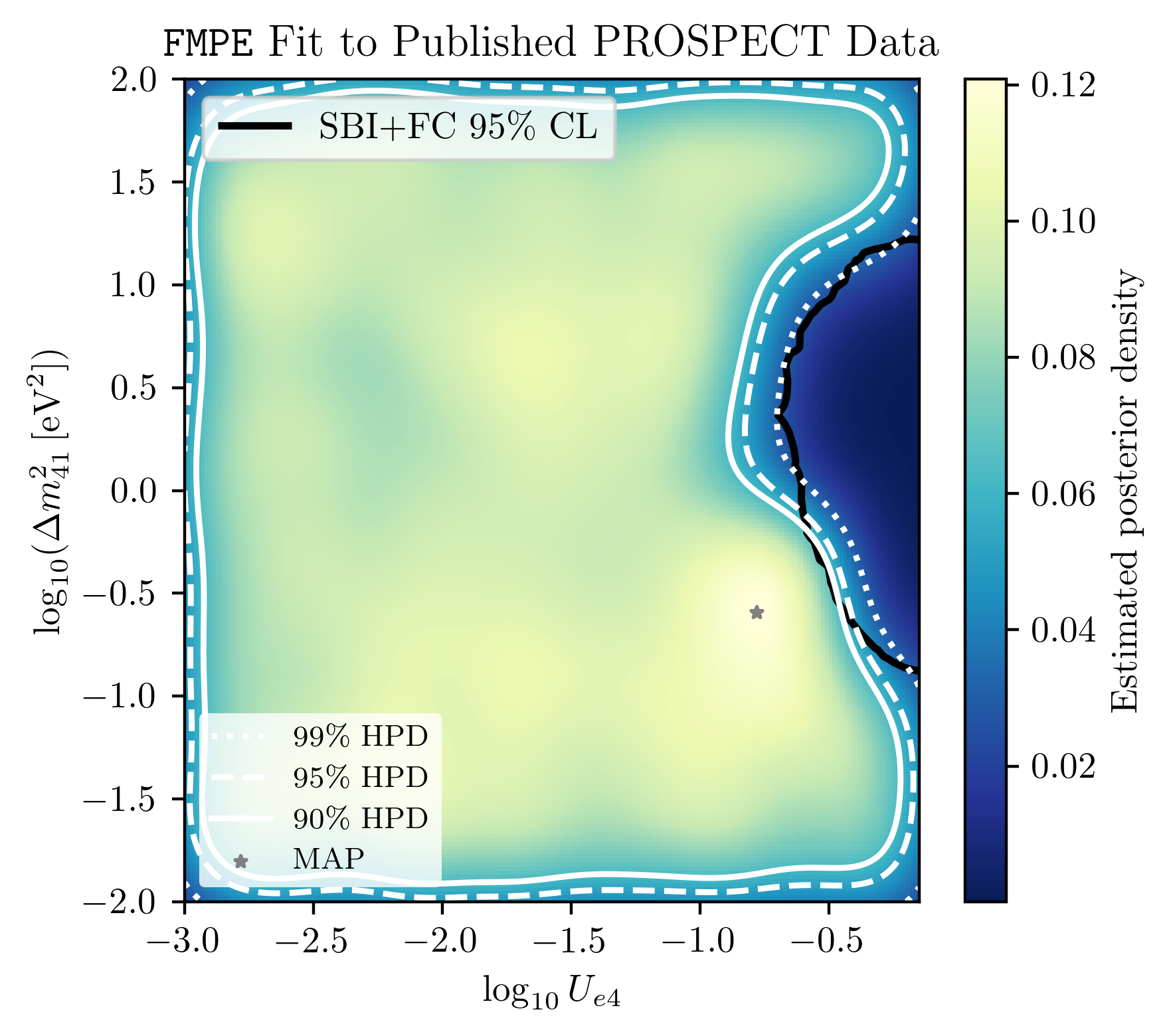}
    \caption{PROSPECT}
\end{subfigure}
\begin{subfigure}[b]{0.32\textwidth}
    \includegraphics[width=\linewidth]{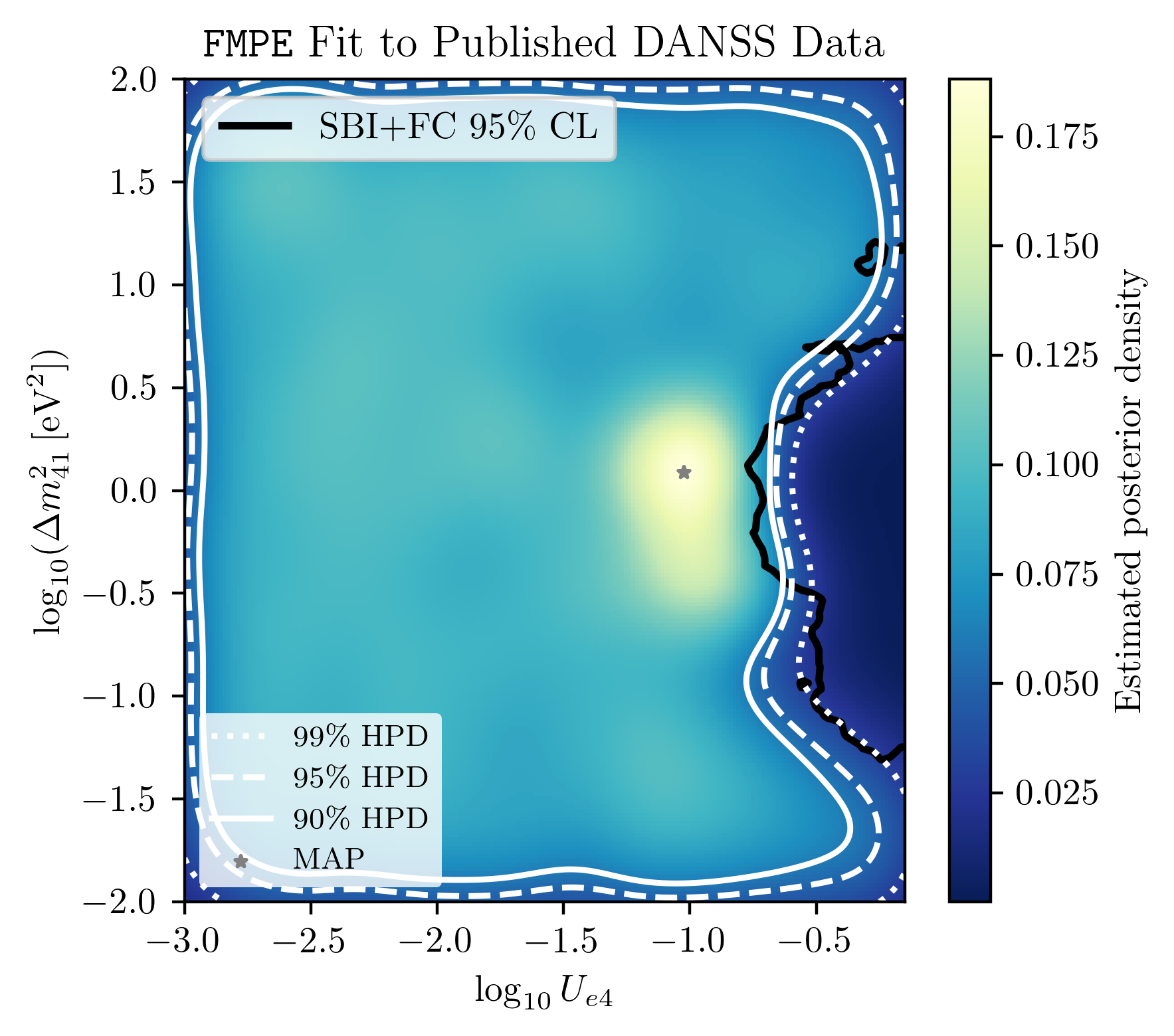}
    \caption{DANSS}
\end{subfigure}
\vspace{0.7em}

\begin{subfigure}[b]{0.32\textwidth}
    \includegraphics[width=\linewidth]{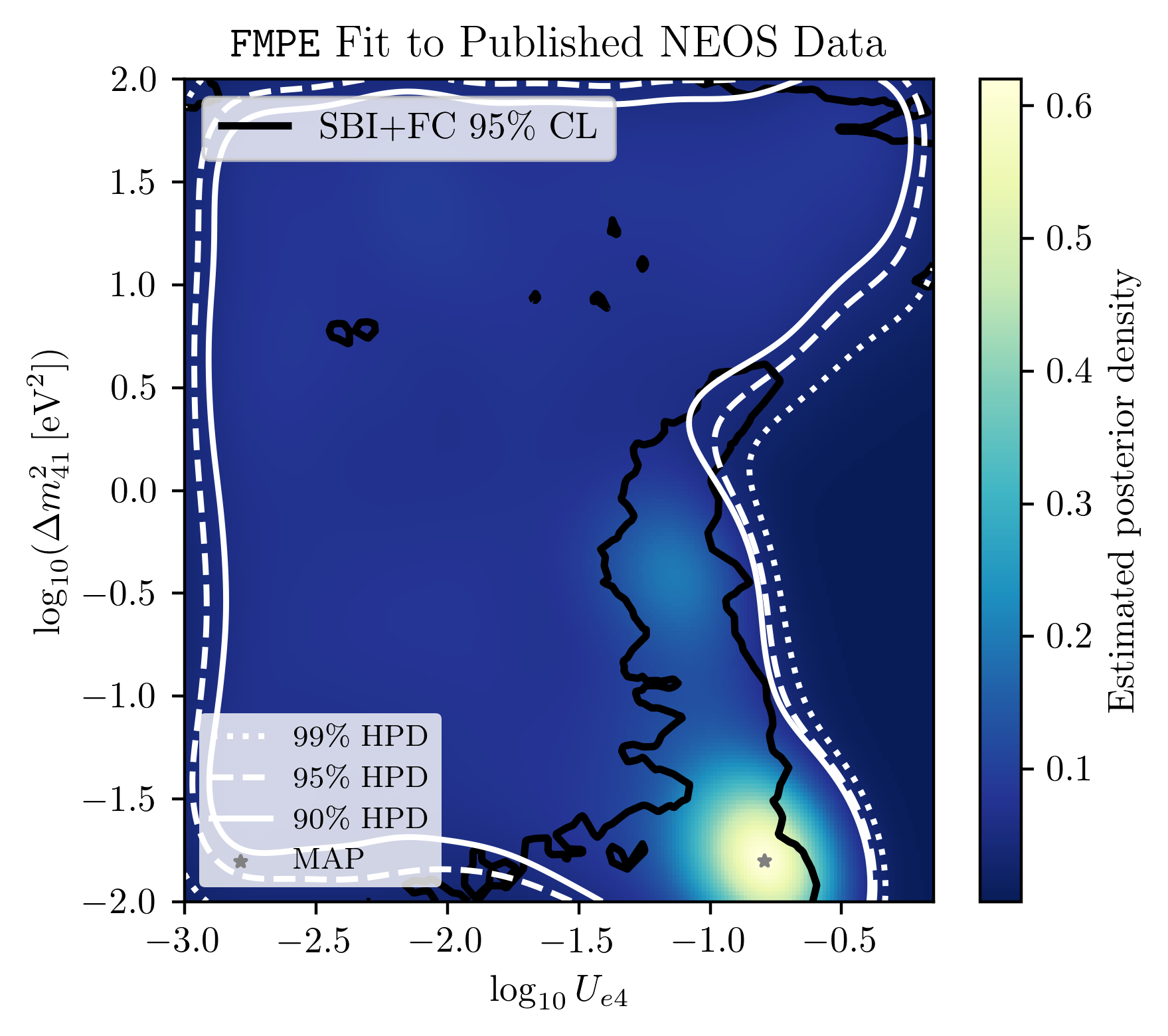}
    \caption{NEOS}
\end{subfigure}
\begin{subfigure}[b]{0.32\textwidth}
    \includegraphics[width=\linewidth]{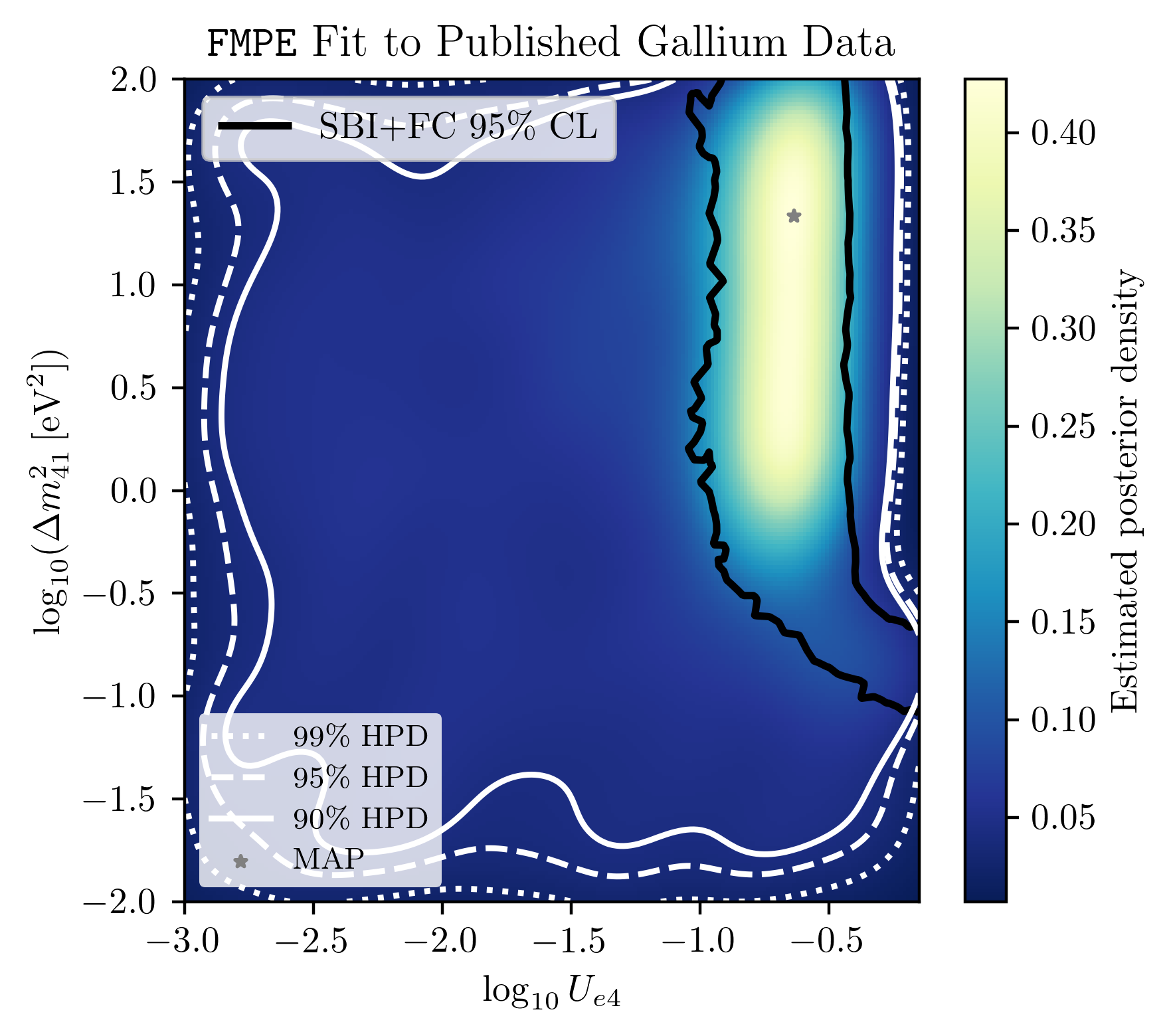}
    \caption{SAGE/GALLEX}
\end{subfigure}
\begin{subfigure}[b]{0.32\textwidth}
    \includegraphics[width=\linewidth]{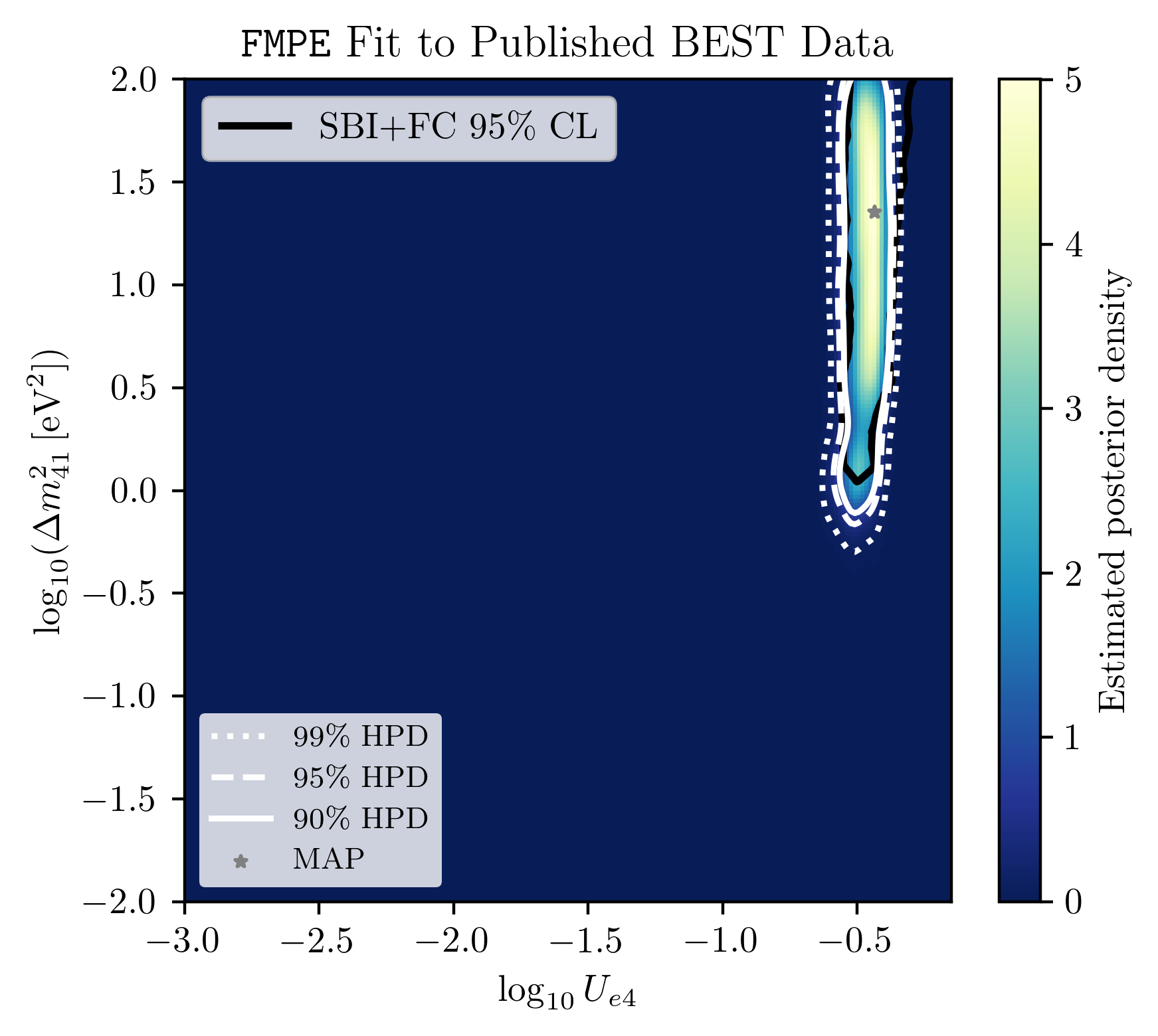}
    \caption{BEST}
\end{subfigure}
\vspace{0.7em}

\begin{subfigure}[b]{0.32\textwidth}
    \includegraphics[width=\linewidth]{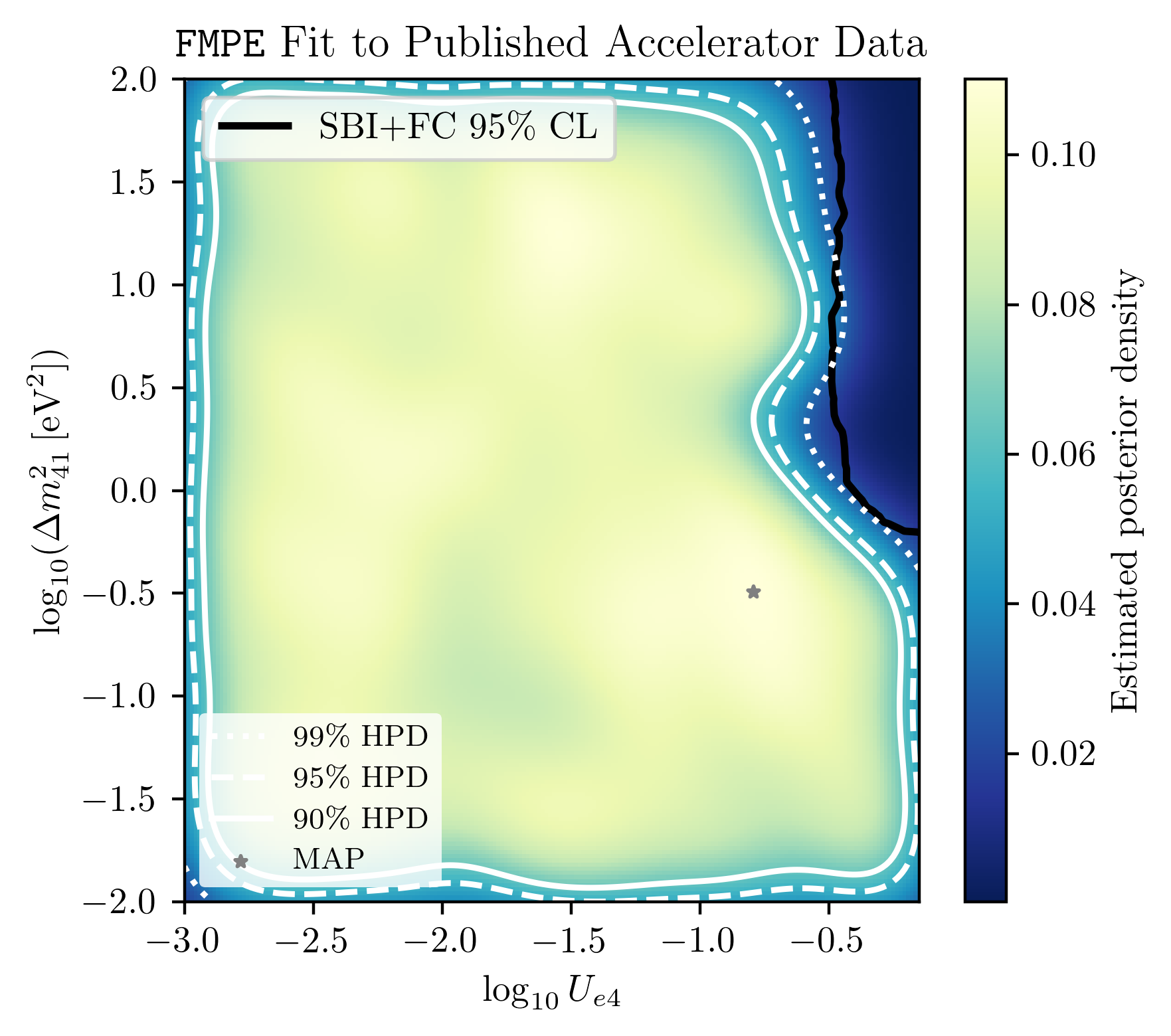}
    \caption{KARMEN/LSND cross section}
\end{subfigure}

\caption{\label{fig:single-experiment-fits}Approximate posterior distributions and highest probability regions using \texttt{FMPE} for each of the single experiments, as well as the combined KARMEN/LSND cross section measurement. Results from the trials-based \texttt{DNRE} construction of frequentist confidence intervals, from Ref.~\cite{10.1088/2632-2153/ae040c}, are also overlaid.}
\end{figure*}

Comparisons of the frequentist trials-based SBI approach (from Ref.~\cite{10.1088/2632-2153/ae040c}) to the posterior estimates from flow matching are shown in Fig.~\ref{fig:single-experiment-fits}.

\clearpage
\clearpage
\bibliography{bibliography}

\end{document}